\documentclass[10pt,conference]{IEEEtran}
\IEEEoverridecommandlockouts
\usepackage{cite}
\usepackage{balance}
\usepackage{amsmath,amssymb,amsfonts}
\usepackage[most]{tcolorbox}
\usepackage{times}
\usepackage{makecell}  
\usepackage{algpseudocode}
\usepackage{graphicx}
\usepackage{textcomp}
\usepackage{marvosym}
\usepackage{xcolor}
\usepackage{multirow} 
\usepackage{multicol} 
\usepackage{color} 
\usepackage{graphicx}   
\usepackage{subcaption} 
\usepackage{float}     
\usepackage{caption}   
\usepackage{subcaption} 
\usepackage{enumitem} 
\usepackage{algorithm} 
\usepackage{array}
\usepackage{xcolor} 
\usepackage{algpseudocode}
\usepackage{longtable}
\usepackage{listings}    
\usepackage{courier}     
\usepackage{subfiles} 
\usepackage{ifsym}
\usepackage{listings}
\usepackage{xcolor}

\usepackage{listings}

\usepackage{setspace}
\usepackage[bottom]{footmisc} 

\tcbset{
  colback=white,
  colframe=black,
  fonttitle=\bfseries,
  coltitle=black,
  boxrule=0.5pt,
  arc=3pt,
  boxsep=3pt,
  left=4pt,
  right=4pt,
  top=4pt,
  bottom=4pt,
  enhanced,
}

\newcommand{\edit}[1]{\textcolor{black}{#1}}

\newcommand{\editgroup}[1]{%
  {\color{black}#1}%
}

\newcolumntype{P}[1]{>{\raggedright\arraybackslash}p{#1}}

\def\BibTeX{{\rm B\kern-.05em{\sc i\kern-.025em b}\kern-.08em
    T\kern-.1667em\lower.7ex\hbox{E}\kern-.125emX}}
\begin{document}

\makeatletter
\renewcommand\footnoterule{%
  \kern-3\p@
  \hrule width 10em height 0.4pt
  \kern2.6\p@
}

\renewcommand{\@makefntext}[1]{%
  \parindent 0em 
  \noindent 
  \hb@xt@1em{\hss\@makefnmark}
  #1
}
\makeatother

\title{EPSO: A Caching-Based Efficient Superoptimizer\\ for BPF Bytecode}





\author{
    Qian Zhu, Yuxuan Liu, Ziyuan Zhu, Shangqing Liu and Lei Bu$^{\dagger}$\thanks{$^{\dagger}$Corresponding author.} \\
	\IEEEauthorblockA{
     \textit{State Key Laboratory of Novel Software Techniques, Nanjing University, Nanjing, Jiangsu 210023, China}\\
     Email: \{zhuqian, yuxuanliu, mg21330081\}@smail.nju.edu.cn, \{shangqingliu, bulei\}@nju.edu.cn
    }\vspace{-8.8mm}

}



\maketitle

\begin{abstract}

Extended Berkeley Packet Filter (eBPF) allows developers to extend Linux kernel functionality without modifying its source code. To ensure system safety, an in-kernel safety checker, the verifier, enforces strict safety constraints (e.g., a limited program size) on eBPF programs loaded into the kernel. These constraints, combined with eBPF’s performance-critical use cases, make effective optimization essential. However, existing compilers (e.g., Clang) offer limited optimization support, and many semantics-preserving transformations are rejected by the verifier, which makes handcrafted optimization rule design both challenging and limited in effectiveness.

Superoptimization overcomes the limitations of rule-based methods by automatically discovering optimal transformations, but its high computational cost limits scalability. To address this, we propose EPSO, a caching-based superoptimizer that discovers rewrite rules via offline superoptimization, and reuses them to achieve high-quality optimizations with minimal runtime overhead.
We evaluate EPSO on benchmarks from the Linux kernel and several eBPF-based projects, including Cilium, Katran, hXDP, Sysdig, Tetragon, and Tracee. \edit{EPSO discovers 795 rewrite rules and achieves up to 68.87\% (avg. 24.37\%) reduction in program size compared to Clang’s output, outperforming the state-of-the-art BPF optimizer K2 on all benchmarks and Merlin on 92.68\% of them.} Additionally, EPSO reduces program runtime by an average of 6.60\%, improving throughput and lowering latency in network applications.



\end{abstract}



\section{Introduction}

Extended Berkeley Packet Filter (eBPF) is a Linux kernel technology that enables safe and high-performance execution of user-defined programs within the kernel, allowing flexible extension of kernel functionality without modifying its source code or loading modules \cite{eBPFDocumentary}. Originally developed for efficient packet filtering in Unix systems \cite{mccanne1993bsd}, eBPF has since been widely adopted in areas such as high-performance networking \cite{hoiland2018express,vieira2020fast,miano2023fast}, observability \cite{scholz2018performance,calavera2019linux,gregg2019bpf,soldani2023ebpf}, and security \cite{security2019blog,wang2022design,nam2022secure}.

eBPF programs, typically written in high-level languages (e.g., C) and compiled to bytecode (e.g., via Clang), are loaded into the kernel after passing a static verifier. The verifier enforces strict safety constraints like size limits, guaranteed termination and aligned memory access. While essential for kernel security, the verifier also introduces challenges. Its size restrictions may exclude well-structured but lengthy programs, and conservative analysis can yield false negatives, rejecting safe programs and hindering efficient eBPF development.
The strict size limitations of eBPF programs, coupled with their increasing adoption in high-performance systems, make effective optimization essential. Optimizing eBPF programs not only allows larger programs to pass the verifier but also improves system throughput and reduces latency in performance-critical tasks. Although modern compilers like Clang provide basic optimizations, they mainly depend on manually crafted rules and consequently miss numerous optimization opportunities. The minimal bytecode size difference between -O2 and -O3 underscores this untapped potential. Moreover, the verifier’s constraints impose additional challenges, making optimization rule design more labor-intensive and less effective.


Recent efforts in eBPF optimization can be broadly classified into three categories: synthesis-based, rule-based, and domain-specific approaches. Synthesis-based methods, exemplified by K2~\cite{xu2021synthesizing}, utilize superoptimization to identify more efficient rewrites.
However, K2 depends on stochastic search, which may overlook valid optimizations. Moreover, its high synthesis latency significantly limits practicality for large-scale eBPF programs. 
Rule-based approaches~\cite{mao2024merlin} employ manually crafted optimization rules, enabling fast and lightweight transformations. Nevertheless, they demand considerable engineering effort to maintain verifier compliance and often fail to generalize beyond predefined rewrite rules, constraining optimization quality. Domain-specific techniques~\cite{bonola2022faster,kuo2022verified} focus on specialized scenarios, such as substituting target program regions with optimized implementations or merging chains of pre-verified programs. While effective within their domains, these methods do not directly optimize eBPF bytecode and lack broad applicability to general-purpose workloads.

To address these challenges, we propose EPSO, a caching-based superoptimizer that efficiently delivers high-quality optimizations for BPF bytecode. Inspired by prior works~\cite{bansal2006automatic,liu2024minotaur}, EPSO offloads the expensive superoptimization process to an offline phase, where it enumeratively explores the program space to collect rewrite rules. These rules are cached and later applied to real-world eBPF code optimization through efficient rule matching, combining the high optimization quality of synthesis-based approaches with the low runtime overhead of rule-based methods. To enhance rule generality and applicability, EPSO introduces a tailored abstraction and slicing strategy: abstraction merges semantically similar rewrites into unified representations, while slicing extracts relevant instructions into self-contained units. Together, these techniques significantly enhance the generality of rewrite rules and improve optimization performance during the rule matching phase.

\begin{figure*}[t]
  \centering
  \includegraphics[width=\textwidth]{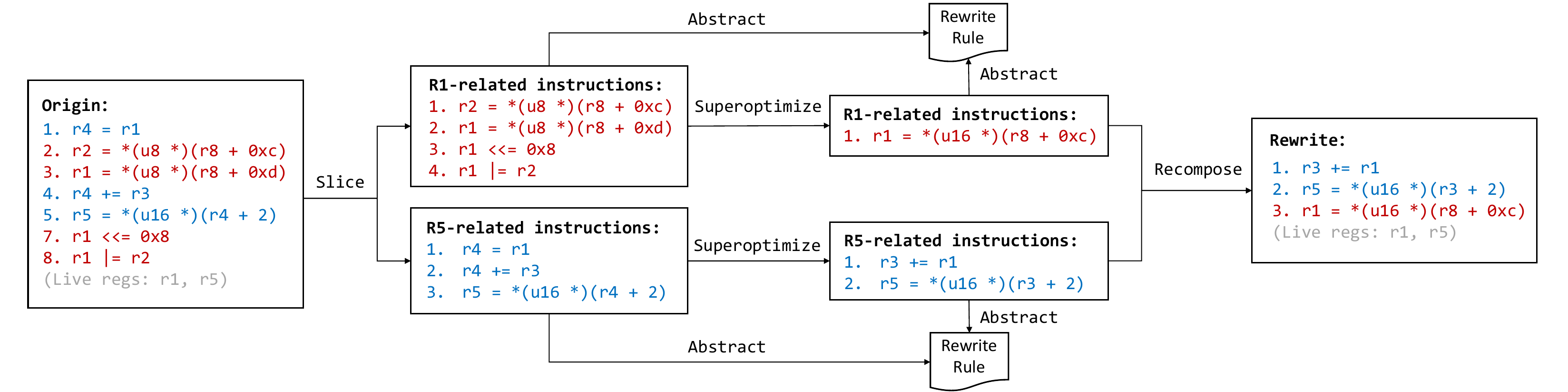}
  \caption{A demonstrative example of EPSO's workflow.}
  \label{fig:motivating_example}
\end{figure*}

We evaluate EPSO on a diverse benchmark set from the Linux kernel \cite{linux_bpf_samples} and several eBPF-based projects, including Cilium \cite{cilium}, Katran \cite{katran}, hXDP \cite{brunella2022hxdp}, Sysdig \cite{sysdig}, Tetragon \cite{tetragon}, and Tracee \cite{tracee}. \edit{EPSO identifies and applies 795 rewrite rules, achieving up to 68.87\% (average 24.37\%) program size reduction compared to Clang’s output, outperforming the state-of-the-art BPF optimizer K2 on all benchmarks and Merlin on 92.68\% of them.} For certain network-oriented programs, EPSO reduces runtime by an average of 6.60\%, improving throughput and latency. Importantly, all optimized programs meet safety constraints and pass the BPF verifier.


In summary, our contributions are as follows:
\begin{itemize}[leftmargin=*]
    \item A novel optimization framework for BPF bytecode that achieves high-quality optimizations efficiently. By offloading superoptimization offline, we precompute a repository of high-quality rewrite rules for efficient reuse during real-world eBPF bytecode optimization. To further enhance rule generality, we introduce novel instruction abstraction and slicing strategies, significantly improving the reusability and applicability of the discovered rewrite rules.
    \item A superoptimization method that synthesizes compact, efficient, and safe rewrite rules. Unlike prior approaches that rely on stochastic search~\cite{xu2021synthesizing}, our method integrates enumerative search with a set of carefully designed pruning strategies, enabling the efficient discovery of a broader set of optimizations during the rewrite rule generation phase.
    \item 
    A comprehensive evaluation on a broad set of benchmarks from diverse sources and scales shows that \edit{our approach achieves up to 68.87\% (avg. 24.37\%) program size reduction}, and up to 19.05\% (avg. 6.60\%) program efficiency improvement. Additionally, it reduces optimization overhead by 88.71\% compared to K2 in synthesis-only mode and incurs negligible overhead when using rule matching.
\end{itemize}


\section{Background and Motivation}

\subsection{Extended Berkeley Packet Filter (eBPF)}

eBPF enhances kernel scalability by enabling sandboxed,
event-driven programs to execute within the kernel, without
requiring kernel source code modifications.
These programs are typically written in high-level languages (e.g., C), compiled to bytecode, and loaded via the \texttt{bpf()} syscall \cite{eBPFDocumentary}. 

Before
execution, an in-kernel safety checker, the verifier, statically
analyzes each program to enforce safety constraints, ensuring
termination, bounded program size, and memory access safety,
etc. Once an eBPF program passes the verifier, it can be
attached to various system events, such as packet reception~\cite{vieira2020fast,brunella2022hxdp}, system calls~\cite{bpftrace}, or application behaviors~\cite{caviglione2021kernel}. Upon triggering, the program executes in the
kernel and communicates with user space through eBPF maps,
enabling efficient and fine-grained system introspection.

\subsection{Motivation}

The strict size constraints of the eBPF verifier and the growing adoption of eBPF in high-performance systems necessitate effective optimization. Optimization enables larger programs to pass the verifier while boosting throughput and lowering latency in critical workloads.
Existing compilers and prior works generally optimize eBPF programs using two approaches: program synthesis and manually designed optimization rules, each with inherent trade-offs. Synthesis-based superoptimizers~\cite{xu2021synthesizing} can discover high-quality optimizations but suffer from high overhead, limiting their scalability. 
Rule-based methods~\cite{mao2024merlin} offer fast and predictable optimizations but often miss optimization opportunities. In the eBPF context, crafting correct and effective rules is especially challenging, as transformations must preserve semantics and comply with the kernel verifier’s safety constraints. 
These limitations reduce the flexibility and effectiveness of rule-based approaches.

To address these challenges, we seek a solution that combines the strengths of synthesis-based and rule-based methods, achieving high-quality optimizations without incurring excessive computational or manual effort. Consequently, we propose a novel BPF bytecode optimization method that uses superoptimization to generate correct, verifier-safe rewrite rules, which are then applied to optimize real-world BPF programs efficiently. To maximize rule generality, we design tailored instruction abstraction and slicing strategies for BPF instructions. This facilitates the construction of a large set of rewrite rules that generalize well and can be applied effectively. Fig.~\ref{fig:motivating_example} illustrates our workflow using a sample instruction sequence. First, the sequence is divided into semantically related segments. Each segment is then independently superoptimized, producing rewrite rules that are abstracted and stored for future reuse. Finally, the optimized segments are recombined, ensuring the original semantics are preserved.
\begin{figure*}[t]
  \centering
  \includegraphics[width=0.8\textwidth]{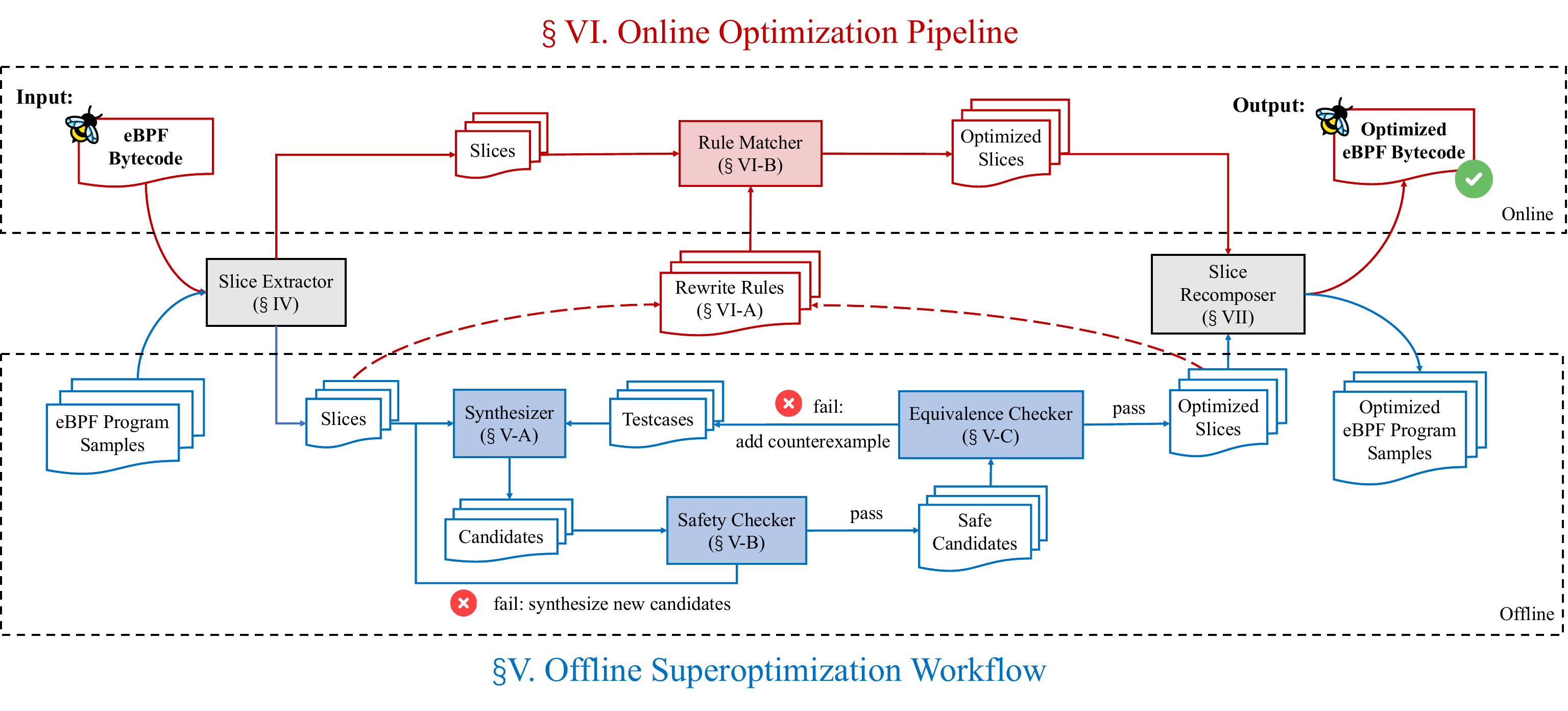}
  \caption{\edit{An overview of EPSO.}}
  \label{fig:flowchartOfEPSO}
\end{figure*}

\section{Overview}

Fig.~\ref{fig:flowchartOfEPSO} illustrates the overall process of EPSO, comprising an offline superoptimization workflow and an online optimization pipeline. The offline phase employs superoptimization to synthesize rewrite rules from representative eBPF sample programs, which are subsequently applied in the online phase to optimize real-world eBPF bytecode efficiently.

In both offline and online phases, inputs are first processed by the \textbf{Slice Extractor}~(\S\ref{sec:slice_extraction}), while the resulting optimized slices are handled by the \textbf{Slice Recomposer} (\S\ref{sec:slice_recomposition}). The extractor partitions instruction sequences into smaller slices that serve as optimization units, and the recomposer reassembles these slices to preserve semantic integrity. 

The offline phase proceeds as follows. Given an instruction slice, the \textbf{Synthesizer} (\S\ref{sec:synthesis}) conducts a cost-guided enumerative search to identify a candidate program that replicates the original’s behavior on a set of input-output testcases. The candidate is first evaluated by the \textbf{Safety Checker} (\S\ref{sec:safety_check}); if it fails to meet verifier constraints, the search continues. Otherwise, it advances to the \textbf{Equivalence Checker} (\S\ref{sec:equivalence_check}) for semantic validation. If the candidate is not equivalent with the original, a counterexample is generated and added to the testcase set to guide subsequent synthesis iterations. Only candidates that pass both safety and equivalence checks are retained. These verified candidates are then paired with their original slices to form \textbf{Rewrite Rules} (\S\ref{sec:representation}).

Once a substantial set of rewrite rules has been collected from representative eBPF sample programs, the online optimizer can be employed to optimize real-world eBPF instruction slices. The online phase employs the \textbf{Rule Matcher} (\S\ref{sec:matching}) to apply offline-synthesized rewrite rules, rewriting original slices into more efficient equivalents.

\section{Extracting Instruction Slices}
\label{sec:slice_extraction}

When optimizing instruction sequences extracted from BPF bytecode, using a sliding window to form instruction slices may group unrelated instructions into the same optimization unit, reducing the reusability of the resulting rewrite rules. To enhance the generality and applicability of rewrite rules, we design a specialized slicing strategy that extracts instruction slices by individually tracing the computation chains of each register or memory region. This enables the precise grouping of optimization-relevant instructions into semantically coherent units, which then serve as reasonable optimization units.

Algorithm~\ref{alg:slice_extraction} illustrates our instruction slicing strategy, which takes an instruction sequence as input and outputs extracted instruction slices that function as optimization units. The overall procedure consists of two major phases: slice construction and slice refinement. 

The \textbf{slice construction} phase (Lines 4–22) builds instruction slices by grouping together instructions that compute values for specific registers or memory locations. For each instruction \textit{I} in the input sequence, the algorithm assigns an identifier \textit{ID} representing the destination it writes to. STORE instructions use a memory-based ID encoded from the base register, offset, and access size, while other instructions use the destination register as ID. Instruction \textit{I} is then added to \textit{Insn\_map[ID]}, along with all instructions related to the registers or memory it reads from, ensuring each slice fully captures the computation chain for its destination. \edit{Notably, since we perform only intra-block optimizations, all instruction sequences selected for slicing originate from basic blocks within the BPF program’s control flow graph. Consequently, instructions within each sequence are executed sequentially, and control-flow information is considered at the beginning of each basic block but not between instructions within the same block.}

The \textbf{slice refinement} phase (Lines 23–24) further processes the collected slices to enhance optimization potential and manageability. It first merges slices related to STORE operations targeting adjacent memory regions, preserving opportunities for memory access fusion that might otherwise be lost. Then, to ensure tractability, each merged slice is partitioned into smaller segments based on a fixed window size. The resulting slices form the final output \textit{Slices}, serving as the fundamental units for downstream optimization.

\begin{algorithm}[!t]
\small
\caption{Instruction Slice Extraction}
\label{alg:slice_extraction}
\begin{algorithmic}[1]
\State \textbf{Input:} Instruction sequence \textit{Insns}
\State \textbf{Output:} Instruction slices functioning as optimization units \textit{Slices}

\State \textit{Insns\_map} $\gets \emptyset$ \Comment{Mapping from register/memory IDs to related instructions}

\For{each instruction $I$ in $Insns$}
    \State \textit{ID} $\gets -1$
    \If{I is a STORE instruction}
        \State \textit{ID} $\gets$ \textit{(I.offset $<<$ 8) $|$ (I.size() $<<$ 4) $|$ I.dst\_reg}
    \Else 
        \State \textit{ID} $\gets I.dst\_reg$
    \EndIf

    \State \textit{Insns\_map[ID]} $\gets I$

    \If{\textit{I} reads from \textit{I.dst
    \_reg}}
        \State  \textit{Insns\_map[ID] += Insns\_map[I.dst\_reg]}
    \EndIf
    \If{\textit{I} reads from \textit{I.src
    \_reg}}
        \State  \textit{Insns\_map[ID] += Insns\_map[I.src\_reg]}
    \EndIf
    \If{I is a LOAD instruction}
        \State \textit{LD\_ID} $\gets$ \textit{(I.offset $<<$ 8) $|$ (I.size() $<<$ 4) $|$ I.src\_reg}
        \State \textit{Insns\_map[ID] += Insns\_map[LD\_ID]}
    \EndIf
\EndFor
\State Merge adjacent memory-related slices in \textit{Insns\_map}.
\State Partition each slice in \textit{Insns\_map} by window size to obtain the final \textit{Slices} as optimization units.
\end{algorithmic}
\end{algorithm}

After slicing, we obtain optimization units which each focuses on the computation of one single register or memory region. This effectively improves the generality of rewrite rules by excluding instructions irrelevant to the optimization target.

\section{Superoptimizing Extracted Slices}
\label{sec:offline}
After slicing instruction sequences into separate slices, we perform superoptimization on these slices to collect rewrite rules. These rules are later applied in Section~\ref{sec:online} to optimize real-world BPF programs.

To guarantee the equivalence between the optimized programs and their originals, superoptimization typically follows the CEGIS~\cite{solar2008program} framework. Within this framework, program equivalence between $P$ and $Q$ is formally defined as:
\editgroup{
\begin{equation}
\label{eq_origin} 
\forall\ \text{input}\ x:\ \text{output}_{P}(x) = \text{output}_{Q}(x) 
\end{equation}
where \(\text{output}_P(x)\) and \(\text{output}_Q(x)\) denote the final runtime states of programs \(P\) and \(Q\) given the same input state \(x\).}

Since directly synthesizing a program equivalent to the original on all inputs is infeasible, CEGIS divides optimization into two stages: candidate synthesis and equivalence verification. In the \textbf{synthesis} (\S\ref{sec:synthesis}) phase, a finite set of testcases (i.e., input-output pairs) serve as the specification, and the synthesizer searches the program space to find a candidate program that produces the same outputs as the original under these testcases. 
During the \textbf{equivalence check} (\S\ref{sec:equivalence_check}) phase, a SMT solver is used to check whether the candidate is strictly equivalent to the original. If equivalent, the candidate is accepted as the final optimized result. If not, a counterexample is generated and added to the testcase set to enhance the constraints of the next synthesis round. As counterexamples accumulate from failed verifications, the constraints are progressively refined, enabling the synthesis-verification loop to converge in a few iterations and ultimately produce an equivalent candidate.

Additionally, to ensure that the optimized program not only preserves semantics but also adheres to the safety constraints enforced by the in-kernel verifier, we introduce a \textbf{safety check} (\S\ref{sec:safety_check}) step prior to equivalence verification. This step enforces a set of safety rules that may otherwise be violated in certain optimization scenarios.



\subsection{Program Synthesis Using Enumerative Search}
\label{sec:synthesis}

As illustrated above, the synthesis stage aims to generate candidate programs that preserve the original program’s behavior across a given set of testcases. Unlike K2, which uses stochastic search that can miss valid optimizations, our approach uses enumerative search~\cite{massalin1987superoptimizer,granlund1992eliminating,bansal2006automatic,phothilimthana2016scaling,alur2017scaling,korf1985depth}. This search strategy systematically explores the program space to find the optimal rewrite equivalent to the original. Furthermore, our method incorporates carefully designed pruning techniques into the search process, substantially reducing the search space and computational overhead to enable more efficient and scalable synthesis. The complete search procedure and pruning strategies are detailed below.


\subsubsection{IDDFS-based Search Strategy}
\label{sec:iddfs}
During the search for candidate programs, we prioritize those with fewer instructions or shorter execution time, enabling us to quickly return the first candidate that meets the test case constraints. To achieve the above search order, we employ Iterative Deepening Depth-First Search (IDDFS)~\cite{korf1985depth} as the primary search algorithm.

IDDFS begins with a shallow depth limit (initially one instruction) and incrementally increases it until a valid candidate is found. This approach ensures that lower-cost candidates are explored first while effectively limiting the search depth.

\subsubsection{Pruning Strategies for Speeding Up Search}
\label{sec:pruning}

Exhaustively exploring the entire program space is time-consuming, making effective pruning essential. A simple yet effective heuristic is that an optimized sequence should never exceed the cost of the original program, and candidates that surpass this cost can be pruned early.

Beyond this, we apply additional pruning strategies specifically designed for the eBPF instruction set. The pruning methods used during the search are as follows:



\textbf{1. Estimating the \textit{distance} to the terminal state.}

To assess the feasibility of a search branch reaching the target terminal state, we define a \textit{distance} metric: the minimum number of instructions needed for the transition from the current state to the target. The formal definition is as follows.

\editgroup{Let the current state $\mathcal{S}$ be represented as $\mathcal{S} = (\mathcal{R}, \mathcal{M})$, where $\mathcal{R} = \{r_i\}_{i=1}^{n}$ denotes the states of registers and $\mathcal{M} = \{m_j\}_{j=1}^{k}$ denotes the states of memory regions. Similarly, let the target state $\mathcal{T} = (\mathcal{R}', \mathcal{M}')$, where $\mathcal{R}' = \{r'_i\}_{i=1}^{n}$ and $\mathcal{M}' = \{m'_j\}_{j=1}^{k}$. We define the \textit{distance} metric $d(\mathcal{S}, \mathcal{T})$ as:

\begin{equation}
d(\mathcal{S}, \mathcal{T}) = \min_{\mathcal{I}} \{|\mathcal{I}| : \mathcal{S} \xrightarrow{\mathcal{I}} \mathcal{T}\}
\end{equation}
where $\mathcal{I} = \{I_1, I_2, \ldots, I_m\}$ denotes an instruction sequence, $|\mathcal{I}|$ represents its length, and $\mathcal{S} \xrightarrow{\mathcal{I}} \mathcal{T}$ indicates that applying the instructions $\mathcal{I}$ to state $\mathcal{S}$ results in state $\mathcal{T}$.

Since modifying a register or memory region requires at least one instruction, $d(\mathcal{S}, \mathcal{T})$ is calculated by counting all differences between the current and target register and memory region states:
\begin{equation}
d(\mathcal{S}, \mathcal{T}) = \sum_{i=1}^{n} \mathbf{1}[r_i \neq r'_i] + \sum_{j=1}^{k} \mathbf{1}[m_j \neq m'_j]
\end{equation}
where $\mathbf{1}[\cdot]$ is the indicator function that equals 1 when the condition is true and 0 otherwise.}

Assuming the optimized sequence does not exceed the original length, we prune any branch whose \textit{distance} exceeds the remaining instruction budget. This heuristic is especially effective for sequences with complex register transformations, where each required state change generally corresponds to at least one instruction, enabling accurate \textit{distance} estimation.

\begin{figure}[!t]
    \centering
    \begin{subfigure}{0.22\textwidth}
        \includegraphics[width=\linewidth]{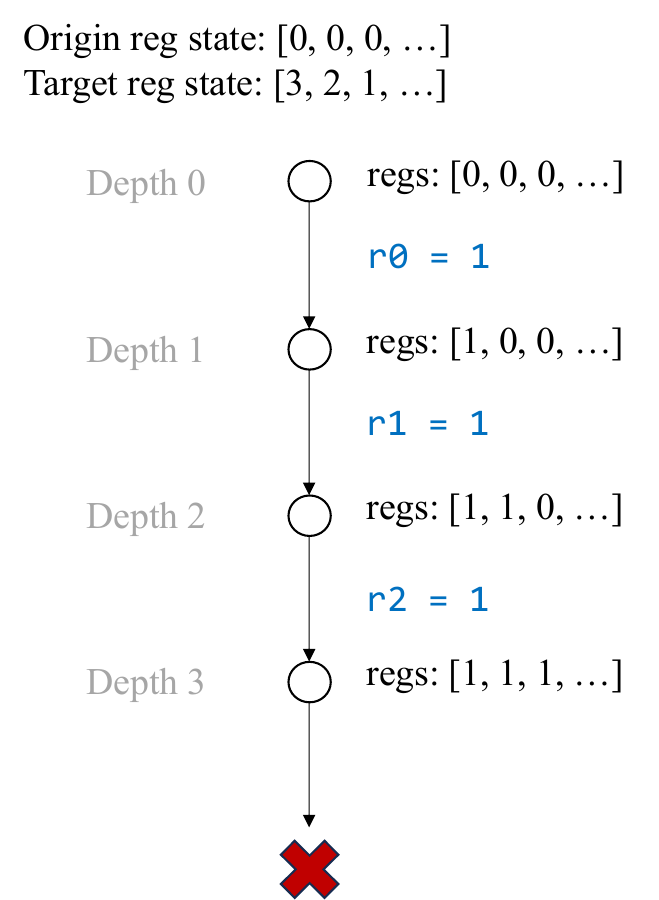}
        \caption{\footnotesize Without distance-estimation.}
        \label{fig:without_dis_est}
    \end{subfigure}
    \hfill
    \begin{subfigure}{0.215\textwidth}
        \includegraphics[width=\linewidth]{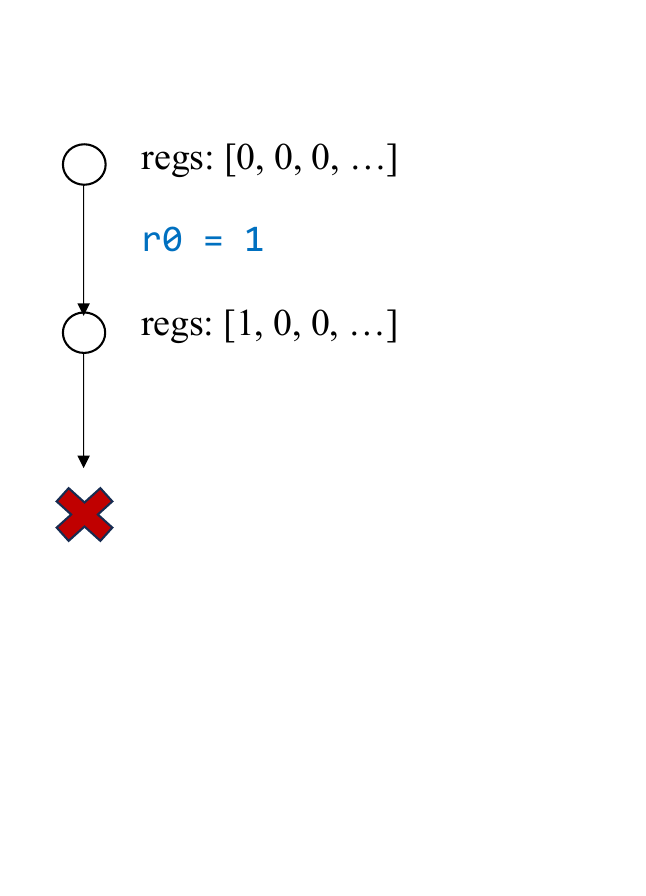}
        \caption{\footnotesize With distance estimation.}
        \label{fig:with_dis_est}
    \end{subfigure}
    \caption{Pruning strategy: distance estimation.}
    \label{fig:pruning2}
\end{figure}

As illustrated in Fig.~\ref{fig:without_dis_est}, without \textit{distance} estimation, the search proceeds until reaching the cost limit. In contrast, Fig.~\ref{fig:with_dis_est} shows that \textit{distance} estimation allows for early termination right after the first invalid instruction. In this example, transforming three registers requires at least three instructions. Once a new instruction fails to reduce the estimated distance, the remaining budget is insufficient and the branch is immediately pruned, preventing unnecessary exploration.



\textbf{2. Memorizing failed explored states.}

\begin{figure}[!t]
    \centering
    \begin{subfigure}{0.45\textwidth}
        \includegraphics[width=\linewidth]{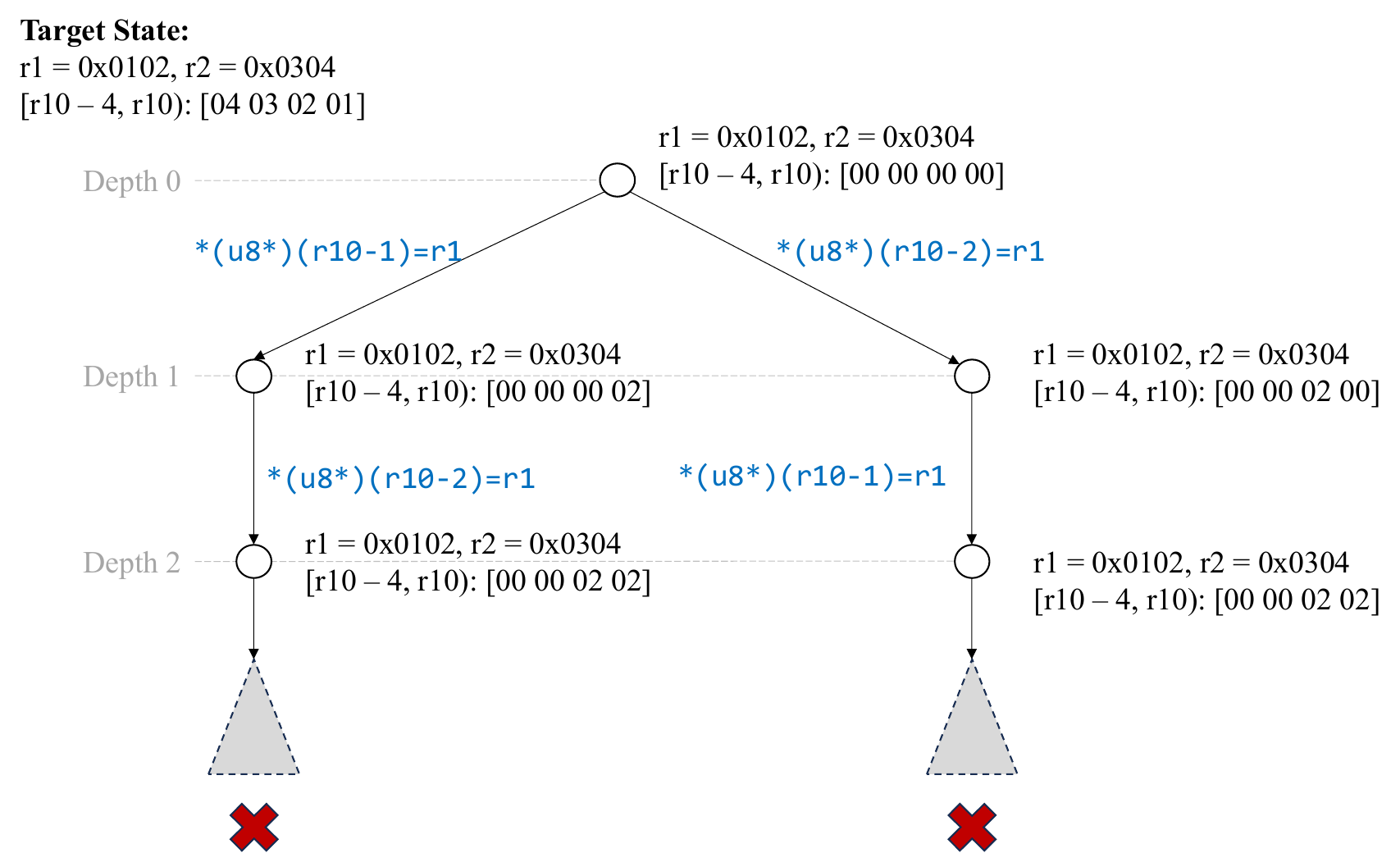}
        \caption{\footnotesize Without failed state memorization.}
        \label{fig:without_fail_mem}
    \end{subfigure}
    \hfill
    \begin{subfigure}{0.45\textwidth}
        \includegraphics[width=\linewidth]{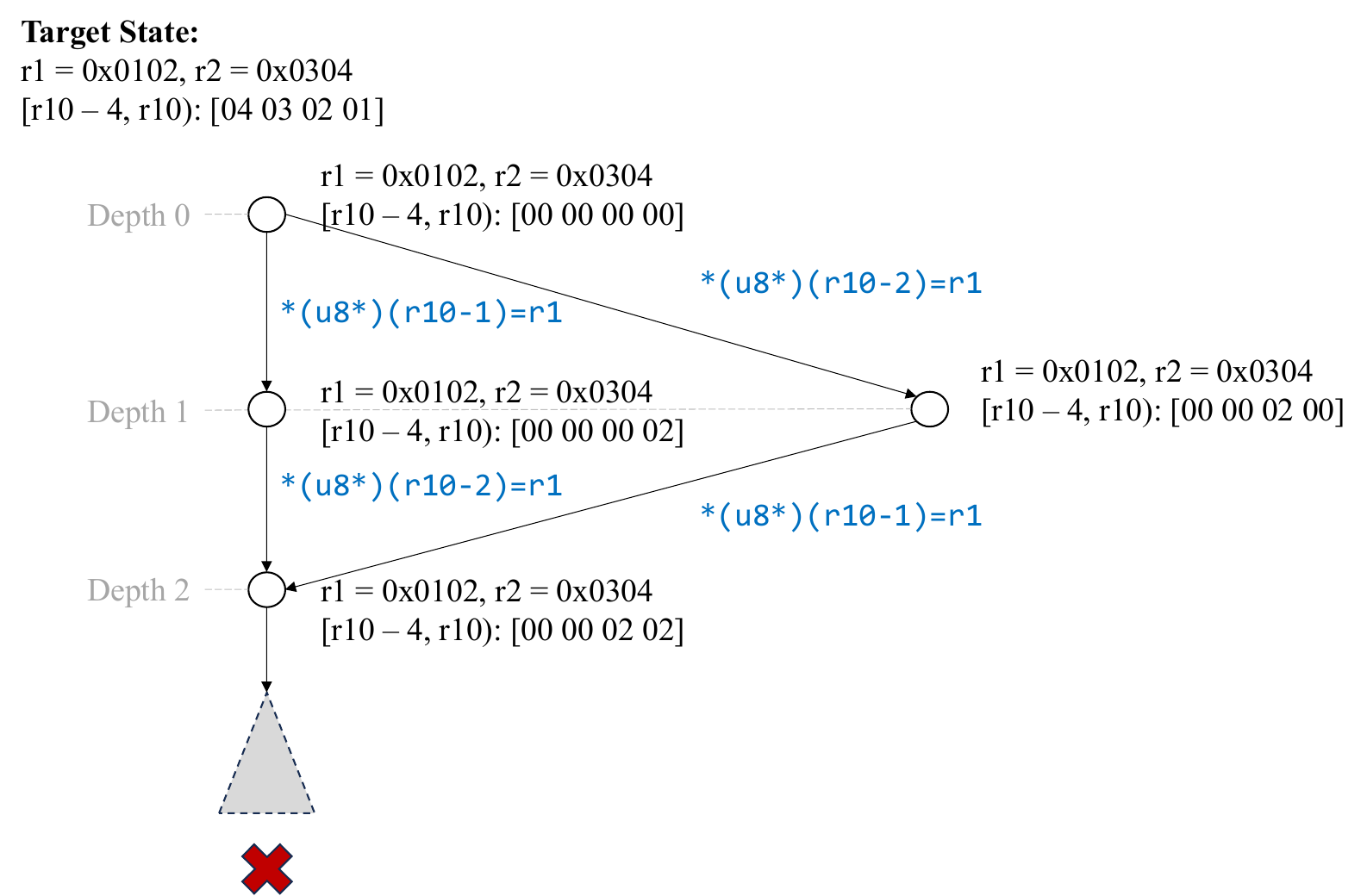}
        \caption{\footnotesize With failed state memorization.}
        \label{fig:with_fail_mem}
    \end{subfigure}
    \caption{Pruning strategy: failed state memorization.}
    \label{fig:pruning3}
\end{figure}

While the first pruning strategy works well for register transformations, it is less effective for memory operations. Memory state differences are harder to quantify due to alignment constraints and uncertain instruction granularity, making distance-based pruning less effective.

To address this, we introduce a memoization-based pruning strategy. During synthesis, we cache program states that led to failed branches along with their failure depths. If the same state recurs at an equal or greater depth, the branch is immediately pruned. Unlike distance-based pruning, this method learns from past failures to avoid redundant exploration.



Consider the following instruction sequence, which stores the lower 16 bits of $r_1$ at memory region $[r_{10} - 2, r_{10})$:

\begin{flushleft}
\small
    \texttt{1. *(u8*)(r10 - 2) = r1} \\
    \texttt{2. r1 >>= 1} \\
    \texttt{3. *(u8*)(r10 - 1) = r1}
\end{flushleft}

A more compact equivalent is:

\begin{flushleft}
\small
    \texttt{1. *(u16*)(r10 - 2) = r1}
\end{flushleft}

During synthesis, redundant searches often occur when semantically equivalent instruction sequences lead to the same memory state. For example, the sequences
\texttt{*(u8*)(r10 - 1) = r1; *(u8*)(r10 - 2) = r1 }
and
\texttt{*(u8*)(r10 - 2) = r1; *(u8*)(r10 - 1) = r1}
produce identical memory states. If the search following the first sequence fails, the second sequence redundantly re-explores the same state and repeats the failure (Fig.~\ref{fig:without_fail_mem}).

Fig.~\ref{fig:with_fail_mem} shows how failed state memoization eliminates this redundancy. By recording failed states with their failure depths, any future visit to the same state at equal or greater depth is pruned immediately. This drastically reduces repeated computations and improves synthesis efficiency.



\textbf{3. Disallowing Redundant Register Definitions.}

During the search, when choosing a destination register, we check whether any previous definition of that register is still unused. If such an unused definition exists, the current search path is pruned immediately, as sequences with redundant operations cannot be optimal.

For example, consider this sequence:

\begin{flushleft}
\small
\texttt{1. r1 = *(u32 *) (r10 + 0)} \\
\texttt{2. r2 = *(u32 *) (r10 + 4)} \\
\texttt{3. r1 = r2}
\end{flushleft}

At instruction 3, $r1$ is redefined without its previous value being used. A definition-use analysis detects this redundancy, allowing the search to prune this suboptimal path early.

\subsection{Safety Check}
\label{sec:safety_check}
Once a candidate is synthesized, we must ensure it meets the verifier’s constraints, including program size limits, structural requirements, and instruction legality. Since our optimization operates at the intra-block level, structural and size checks are less critical. The primary challenge lies in ensuring the transformed code complies with the verifier’s instruction legality.

To address this, we analyze the verifier’s core instruction validation routine, the \textit{do\_check} function, and encode its relevant constraints into the synthesis process to ensure all generated candidates can pass the verifier.

In practice, we observed several cases where semantically correct candidates were rejected due to subtle verifier rules. Two representative examples are discussed below.



\begin{figure}[!t]
    \centering

    \begin{subfigure}{0.8\linewidth}
        \centering
        \includegraphics[width=0.8\linewidth]{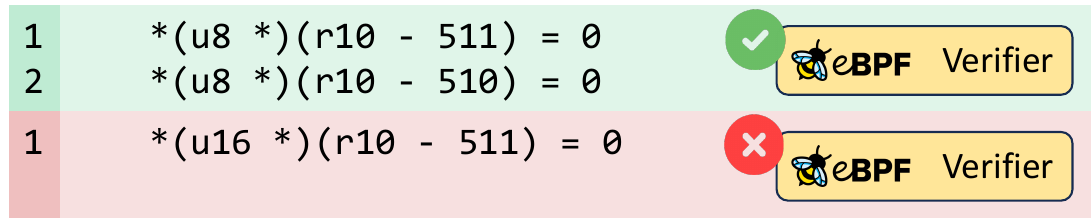}
        \caption{Unsafe optimization due to memory misalignment.}
        \label{fig:verifier_memory_alignment}
    \end{subfigure}

    \vspace{0.5cm}  

    \begin{subfigure}{0.8\linewidth}
        \centering
        \includegraphics[width=0.8\linewidth]{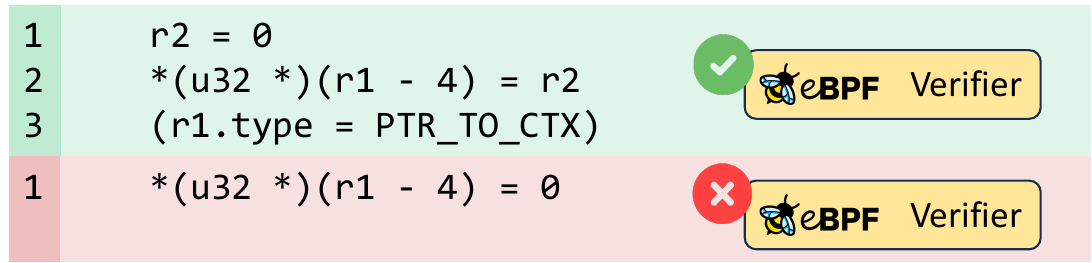}
        \caption{Unsafe Optimization due to pointer type.}
        \label{fig:ptr_to_ctx}
    \end{subfigure}

    \caption{Unsafe optimizations rejected by the verifier.}
    \label{fig:safety_check}
\end{figure}




\textbf{1. Memory address alignment.}


As shown in Fig.~\ref{fig:verifier_memory_alignment}, merging two 8-bit store instructions into a single 16-bit store is a valid optimization in principle. However, the verifier enforces strict memory alignment: the offset from the stack base to the 16-bit store address must be a multiple of 16. As a result, this optimization is invalidated, as it violates the verifier's memory access alignment rule.

\textbf{2. The destination register type of the $BPF\_ST$ instruction cannot be $PTR\_TO\_CTX$.}


As shown in Fig.~\ref{fig:ptr_to_ctx}, although the optimized code is semantically equivalent to the original when register $r2$ is no longer live, the verifier rejects it due to a strict rule: if the base register in a memory access is a context pointer, the source operand must not be an immediate value. 



\subsection{Equivalence Check}
\label{sec:equivalence_check}

After verifying the candidate’s safety, we perform final equivalence checking to ensure it is functionally equivalent to the original. The equivalence between program $P$ and program $Q$ is defined in Eq.~\ref{eq_origin}, and the negation of this formula is:
\editgroup{
\begin{equation}\label{eq_negation} \exists\ \text{input}\ x:\ \text{output}_{P}(x) \neq \text{output}_{Q}(x) \end{equation}
}

By proving Eq. \ref{eq_negation} false, we can establish the equivalence of the two programs. If Eq. \ref{eq_negation} is satisfiable, it indicates that there exists at least one input for which programs $P$ and $Q$ produce different outputs, thereby disproving their equivalence. Conversely, if Eq.~\ref{eq_negation} is unsatisfiable, no such input exists, and the equivalence of $P$ and $Q$ is proven.


Based on Eq.\ref{eq_negation}, we formulate the final constraint for the SMT solver as shown in Eq.\ref{eq:equivalence}. To further expand the optimization space, we incorporate additional constraints into this formula in practice, including register and memory liveness, as well as register value ranges.

\begin{equation}
\begin{aligned}
&\text{input}_P = \text{input}_Q \\
&\land \text{execution behavior of program P} \\
&\land \text{execution behavior of program Q} \\
&\land \text{output}_P \neq \text{output}_Q
\end{aligned}
\label{eq:equivalence}
\end{equation}

\section{Representing and Matching Rewrite Rules}
\label{sec:online}
\subsection{Representing Rewrite Rules}
\label{sec:representation}
After optimizing the extracted instruction slices, we obtain a set of rewrite rules, each comprising a pair of original and rewritten instruction slices. Using concrete operand values to represent these rules leads to overly specific rewrite rules that are difficult to match in practice. In reality, the effectiveness of a rewrite rule largely depends on factors such as instruction opcodes and relative memory offsets, rather than specific register identifiers or accurate memory addresses.

To enhance the generality of rewrite rules, we abstract away non-essential operand values, retaining only those that directly influence the optimization. This abstraction allows the rules to be applied more broadly across diverse instruction instances.

\begin{figure}[!t]
    \centering
    \begin{subfigure}{0.23\textwidth}
        \includegraphics[width=\linewidth]{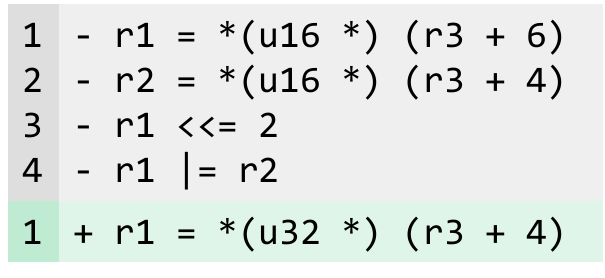}
        \caption{Original rewrite rule}
    \end{subfigure}
    \hfill
    \begin{subfigure}{0.23\textwidth}
        \includegraphics[width=\linewidth]{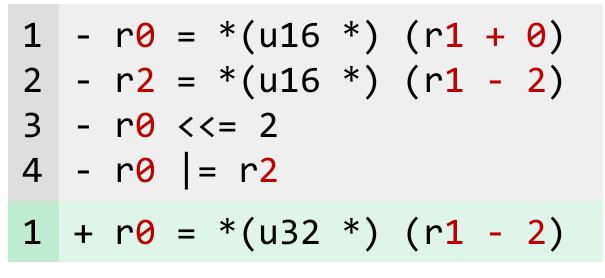}
        \caption{Abstracted rewrite rule}
    \end{subfigure}
    \caption{Example of rewrite rule abstraction.}
    \label{fig:abstraction}
\end{figure}

We implement this abstraction through a normalization process. Registers are renamed in order of appearance using symbolic identifiers (e.g., $r_0$, $r_1$, etc.). For memory offsets, the first offset relative to a base register is normalized to 0, and subsequent offsets from the same base are expressed as relative differences. An example is illustrated in Fig.~\ref{fig:abstraction}.



\subsection{Matching Rewrite Rules}
\label{sec:matching}
Since rewrite rules are stored in abstracted form, the instruction sequence to be optimized must undergo the same abstraction process to enable successful rule matching. After a match is found, a de-abstraction step is performed to reconstruct the concrete optimized sequence. This step leverages the mapping between abstract and real registers, as well as the recorded base offsets, both collected during the original abstraction phase. \edit{Additionally, after each rule application, we perform formal equivalence and safety checks to ensure that all transformations preserve program semantics and safety.}





\section{Recomposing Optimized Instruction Slices}
\label{sec:slice_recomposition}
With extracted instruction slices optimized, we need to further recompose the optimized instruction slices into a complete instruction sequence. To ensure the correctness of the reassembled slices, the recombination must respect the def-use relationships of registers and memory regions.

\begin{algorithm}[!t]
\small
\caption{Instruction Slice Recomposition}
\label{alg:slice_recomposition}
\begin{algorithmic}[1]
\State \textbf{Input:} Optimized instruction slices \textit{Slices}
\State \textbf{Output:} Optimized instruction sequence \textit{Sequence}

\State \textit{Insns\_set} $\gets \emptyset$
\State \textit{In\_edges} $\gets \emptyset$
\State \textit{Out\_edges} $\gets \emptyset$
\For{each slice $(id, \textit{insns})$ in \textit{slices}}
    \For{each instruction \textit{I} in \textit{insns}}
        \State $ID \gets$ definition ID of $I$
        \For{each instruction $J$ in \textit{insns} after \textit{I} }
            \If{$J$ uses $ID$}
                \State $In\_edges[J] += I$
                \State $Out\_edges[I] += J$
            \EndIf
            \If{$J$ redefines $ID$}
                \State \textbf{break}
            \EndIf
        \EndFor

        \If{I is identical to instruction J in \textit{insns\_set}}
            \State Redirect I related edges to J
        \Else
            \For{each instruction $J$ in \textit{insns\_set}}
                \If{$I$ and $J$ come from different slices}
                    \State Add cross-slice dependency edges \State based on def-use relations
                \EndIf
            \EndFor
            \State Add I to \textit{insns\_set}
        \EndIf
    \EndFor
\EndFor

\State \textit{Sequence} $\gets$ \textit{TopologicalSort(in\_edges, out\_edges)}
\State Append unsorted nodes from \textit{insns\_set} to \textit{Sequence}
\State \Return \textit{Sequence}
\end{algorithmic}
\end{algorithm}

Algorithm~\ref{alg:slice_recomposition} presents a def-use-aware instruction slice recomposition strategy. The core idea is to construct a def-use graph of instructions and reorganize instructions from different slices following a topological ordering. The process is structured as follows: 

The \textbf{intra-slice dependency construction} phase (Lines 9–17) analyzes data dependencies for each slice. It tracks each instruction’s defined value (e.g., register or memory location) and adds edges to subsequent instructions that use this value, stopping when a redefinition is encountered.

The \textbf{cross-slice deduplication and dependency construction} phase (Lines 18–28) merges identical instructions across slices to reduce redundancy. To preserve semantics correctness, it also adds cross-slice dependency edges based on def-use relationships.

The \textbf{topological ordering} phase (Lines 31–32) applies topological sort to determine a valid instruction order based on the dependency graph. Unrelated instructions not included in the graph are appended at the end.

\section{Evaluation}

We evaluate the performance of EPSO by answering the following questions:
\begin{itemize}[leftmargin=*]
    \item RQ1: How effectively does EPSO reduce program size to improve compactness? (\S\ref{sec:exp1})
    \item RQ2: How effectively does EPSO improve program efficiency? (\S\ref{sec:exp2})
    \item RQ3: What is the optimization overhead introduced by EPSO, and how do different pruning strategies impact it during the superoptimization phase? (\S\ref{sec:exp3})
    \item RQ4: How effective are the rewrite rules generated during the superoptimization phase? (\S\ref{sec:exp4})
\end{itemize}

\noindent\textbf{Experimental Setup.} We conducted all experiments on a workstation with Intel Core i7-12700H CPU (2.30GHz), 32GB RAM, running Ubuntu 22.04.1 LTS with Linux kernel v5.15.0.

\begin{table*}[htbp]
  \centering
  \caption{\edit{Experimental configurations for each research question (RQ).}}
  \label{tab:rq-configs}
  \resizebox{\linewidth}{!}{%
  \editgroup{
  \begin{tabular}{|c|c|c|c|c|c|}
    \hline
    RQ & Baseline(s) & Compiler Optimization Level & Optimization Mode & Rule-extraction Set & Optimization Set \\
    \hline
    \multirow{3}{*}{\centering RQ1} 
      & K2 & -Os & Offline & -- & K2 benchmark suite  \\
    \cline{2-6}
      & Merlin & -O2 & Offline & -- & Merlin benchmark suite \\
    \cline{2-6}
      & K2 \& Merlin & -O2 & Offline & -- & 30 eBPF programs from diverse sources \\
    \hline
    RQ2 & K2 \& Merlin & -O3 & Offline & -- & 6 performance-oriented eBPF programs \\
    \hline

\multirow{3}{*}{\centering RQ3} 
  & \multirow{3}{*}{K2} 
  & \multirow{3}{*}{-Os} 
  & Offline & -- & K2 benchmark suite \\
\cline{4-6}
  &   &   & Online & \multicolumn{2}{c|}{K2 benchmark suite} \\
\cline{4-6}
  &   &   & Hybrid & \multicolumn{2}{c|}{K2 benchmark suite} \\
\hline

    RQ4 & EPSO in offline mode & -O2 & Online & 20\% of Merlin benchmark suite & 80\% of Merlin benchmark suite \\
    \hline
  \end{tabular}
  }
  }
\end{table*}

\noindent\textbf{Benchmark Selection.}
The benchmark suites used in this paper are sourced from several open-source projects containing eBPF programs: the Linux kernel~\cite{linux_bpf_samples} Cilium~\cite{cilium} Katran~\cite{katran} Tetragon~\cite{tetragon} Tracee~\cite{tracee} Sysdig~\cite{sysdig} and hXDP~\cite{brunella2022hxdp}.

As different RQs focus on different aspects, and existing works such as K2 and Merlin also adopt distinct benchmark selections. To comprehensively compare the performance of our tool EPSO with K2 and Merlin, we select tailored benchmark suites for each RQ. Specifically:

For \textbf{RQ1 (program compactness)}, we use three types of benchmark suites:
\begin{itemize} [leftmargin=*]
    \item K2 benchmark suite: 19 programs selected by K2 as benchmarks, sourced from the Linux kernel~\cite{linux_bpf_samples}, Cilium~\cite{cilium}, Katran~\cite{katran}, and hXDP~\cite{brunella2022hxdp}.
    \item Merlin benchmark suite: 519 eBPF programs selected by Merlin as benchmarks, drawn from Tetragon~\cite{tetragon}, Sysdig~\cite{sysdig}, and Tracee~\cite{tracee}.
    \item Diverse benchmark suite: 30 eBPF programs from various sources, including the Linux kernel~\cite{linux_bpf_samples}, Tetragon~\cite{tetragon}, Cilium~\cite{cilium}, and Katran~\cite{katran}.
\end{itemize}

Due to differences in tool implementation and input format support, Merlin cannot be evaluated on K2’s benchmarks, which are encoded using a special format in K2’s artifact 
(supported by EPSO but not by Merlin, which only accepts IR or bytecode as input).
Conversely, K2 is not evaluated on Merlin’s benchmarks because they are generally large-scale, and K2’s optimization process is prohibitively slow (e.g., nearly 48 hours for a program with over 1700 instructions).

To ensure a fair comparison, we evaluate EPSO against K2 using the K2 benchmark suite and against Merlin using the Merlin suite. Additionally, we introduce a third benchmark suite of 30 diverse eBPF programs spanning multiple sources and program sizes, enabling a direct comparison of EPSO, K2, and Merlin in terms of program size reduction.

For \textbf{RQ2 (program efficiency)}, we select 6 high-performance eBPF programs from the Linux kernel~\cite{linux_bpf_samples}, comparing runtime reductions of EPSO, K2 and Merlin.


For \textbf{RQ3 (optimization overhead)}, we compare EPSO and synthesis-based K2 using the K2 benchmark suite from RQ1. This evaluates their optimization overhead and EPSO’s performance across different configurations. Additionally, we use the first 18 small programs from the suite to study the impact of pruning strategies, as optimization overhead can be excessive for large eBPF programs without pruning.


\editgroup{For \textbf{RQ4 (rewrite rule effectiveness)}, we use 20\% of the Merlin benchmark suite from RQ1 as the rule-extraction set and the remaining 80\% as the optimization set, where the extracted rewrite rules are applied to perform optimizations. The optimization results are compared with direct superoptimization on the same benchmarks, demonstrating the generality and effectiveness of the learned rules.}

\editgroup{
\textbf{Experimental Configurations.} 
Since different RQs focus on distinct evaluation aspects and baselines, their experimental setups differ in compiler optimization level, optimization mode (offline or online), and dataset partitioning for rule-extraction and optimization. Table~\ref{tab:rq-configs} summarizes these configurations.

For RQ1 and RQ2, which evaluate optimization effects, we apply direct superoptimization to all benchmarks. Consequently, as shown in Table~\ref{tab:rq-configs}, the rule-extraction set is empty and the validation set corresponds to the benchmark suite directly optimized through superoptimization. 

For RQ3, which investigates the optimization overhead of EPSO under different configurations, evaluations are conducted in offline, online, and hybrid modes. In offline mode, benchmarks are directly superoptimized. In online mode, rewrite rules are first extracted from the target benchmarks, and then applied in online optimization to measure its overhead. In hybrid mode, the strategy is similar, except that superoptimization is applied when rule matching fails. As the rule-extraction and optimization sets are identical, EPSO achieves the same optimization effects across all three configurations. In RQ4, we will examine differences in optimization effects using distinct rule-extraction and optimization sets.

For RQ4, which examines the transferability of rewrite rules, we evaluate EPSO in the online mode using 20\% of Merlin’s benchmark suite as the rule-extraction set and the remaining 80\% as the optimization set. The results are then compared with those obtained through direct superoptimization, allowing us to assess the effectiveness and generalizability of the collected rules on unseen benchmarks.

Finally, the compiler optimization level is set to align with baseline configurations for fair comparison and to accommodate the specific optimization objectives.

}

\begin{figure*}[t]
    \small
    \centering
    \begin{subfigure}[t]{0.32\textwidth}
        \includegraphics[width=\linewidth]{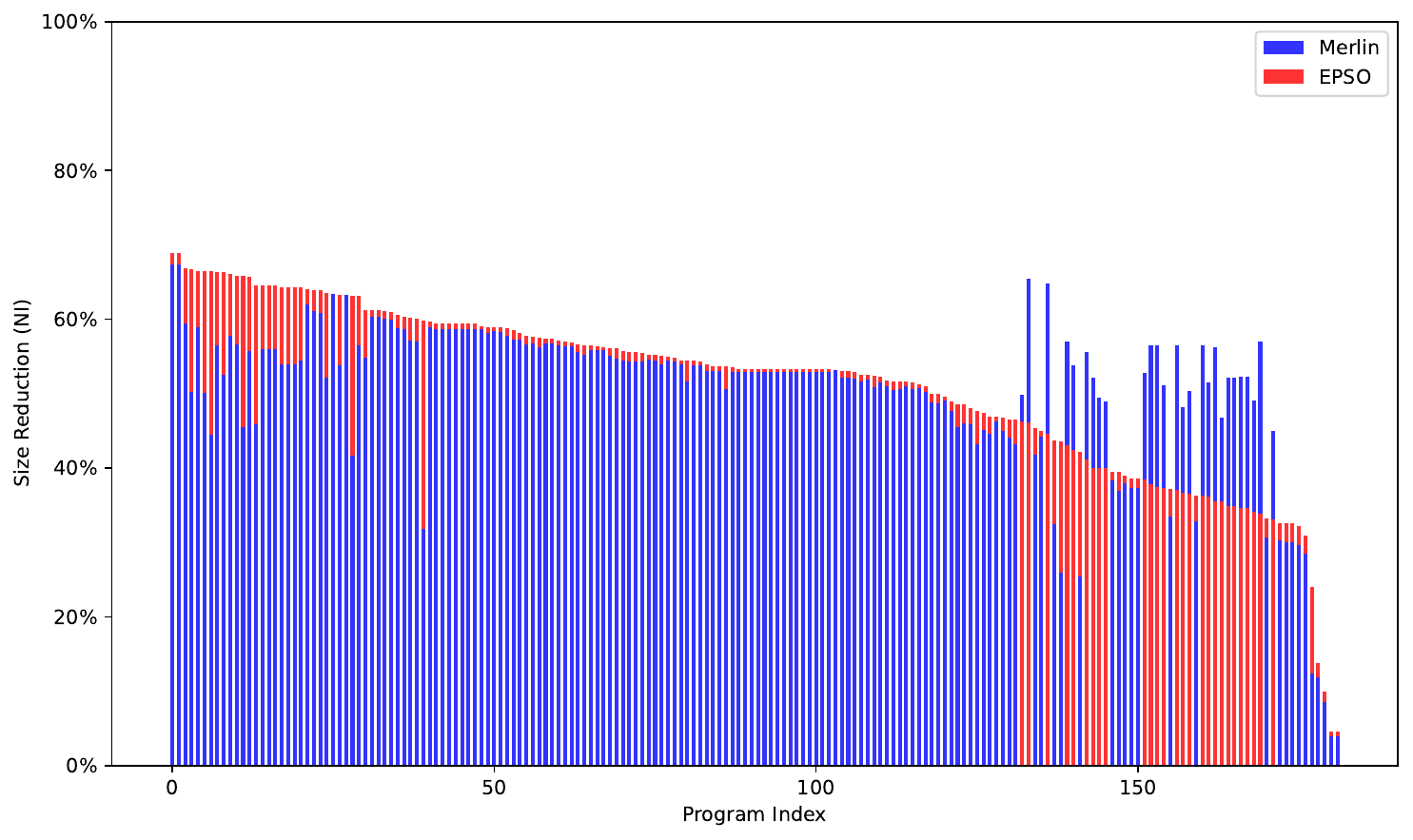}
        \caption{\edit{Compactness comparison between EPSO and Merlin on Sysdig.}}
        \label{fig:sysdig}
    \end{subfigure}
    \hfill
    \begin{subfigure}[t]{0.32\textwidth}
        \includegraphics[width=\linewidth]{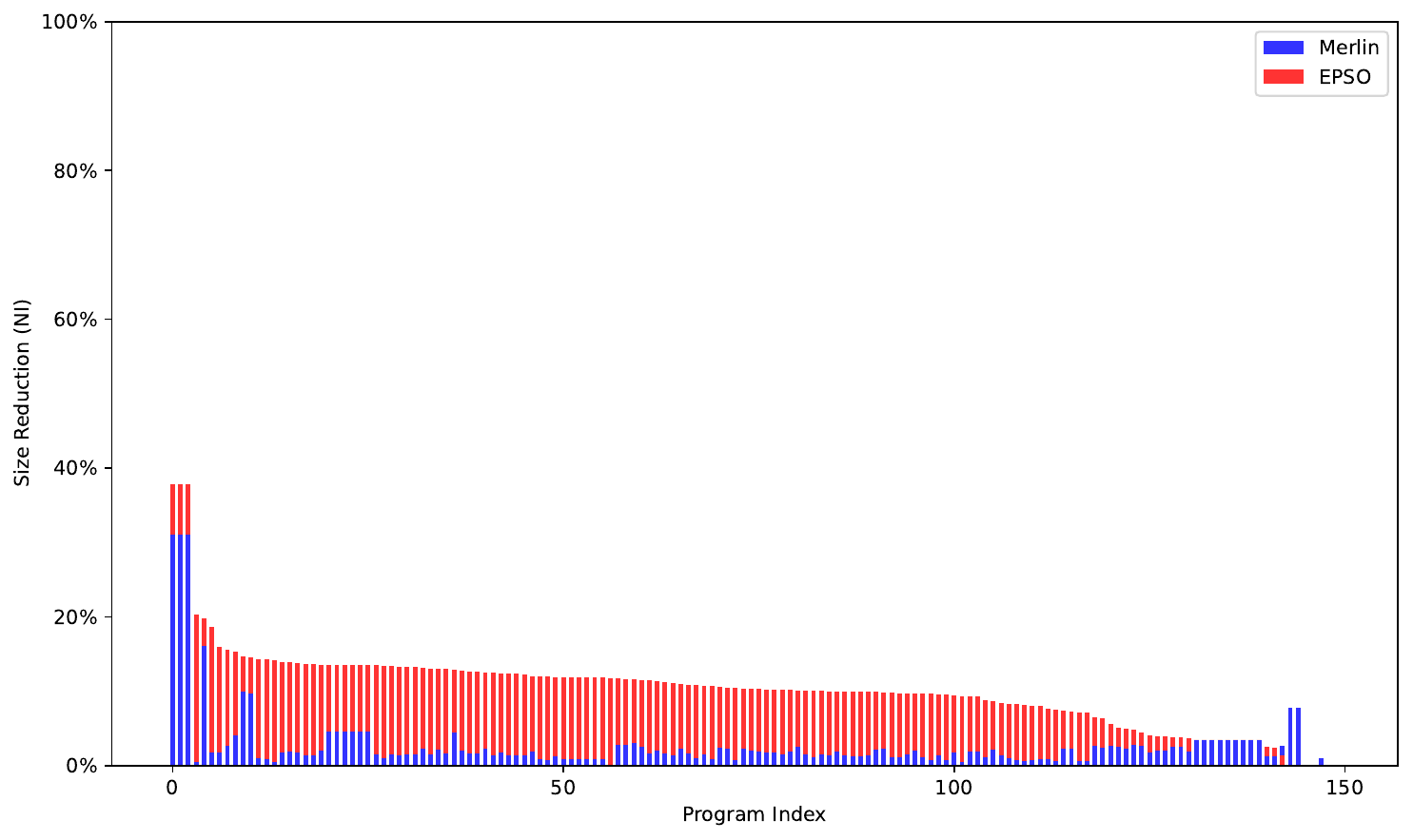}
        \caption{\edit{Compactness comparison between EPSO and Merlin on Tracee.}}
        \label{fig:tracee}
    \end{subfigure}
    \hfill
    \begin{subfigure}[t]{0.32\textwidth}
        \includegraphics[width=\linewidth]{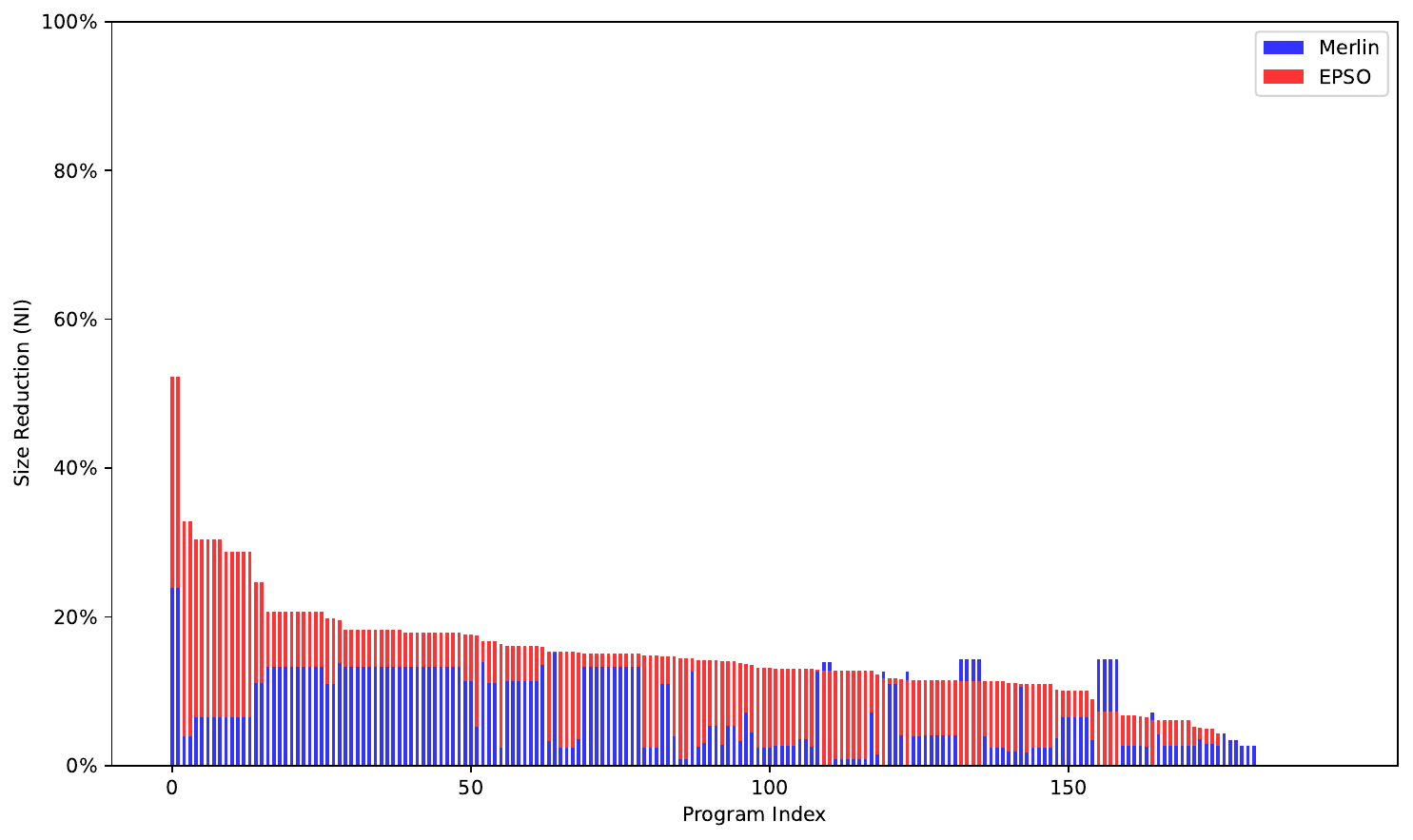}
        \caption{\edit{Compactness comparison between EPSO and Merlin on Tetragon.}}
        \label{fig:tetragon}
    \end{subfigure}
    
    \vspace{1em} 
    
    \begin{subfigure}[!t]{0.43\textwidth}
        \includegraphics[width=\linewidth]{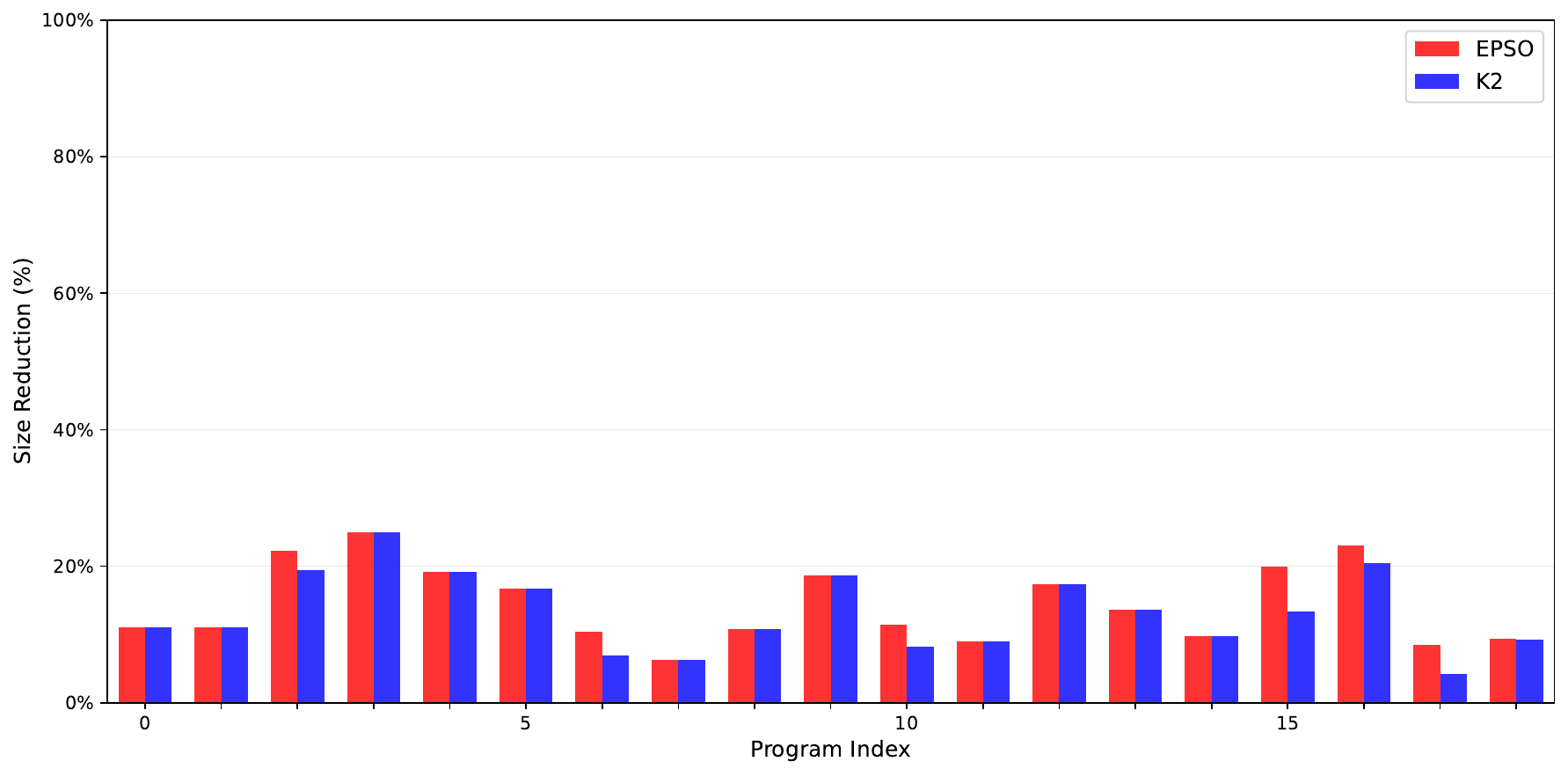}
        \caption{Compactness comparison between EPSO and K2 on the K2 benchmark suite.}
        \label{fig:k2}
    \end{subfigure}
    \hfill
    \begin{subfigure}[!t]{0.50\textwidth}
        \includegraphics[width=\linewidth]{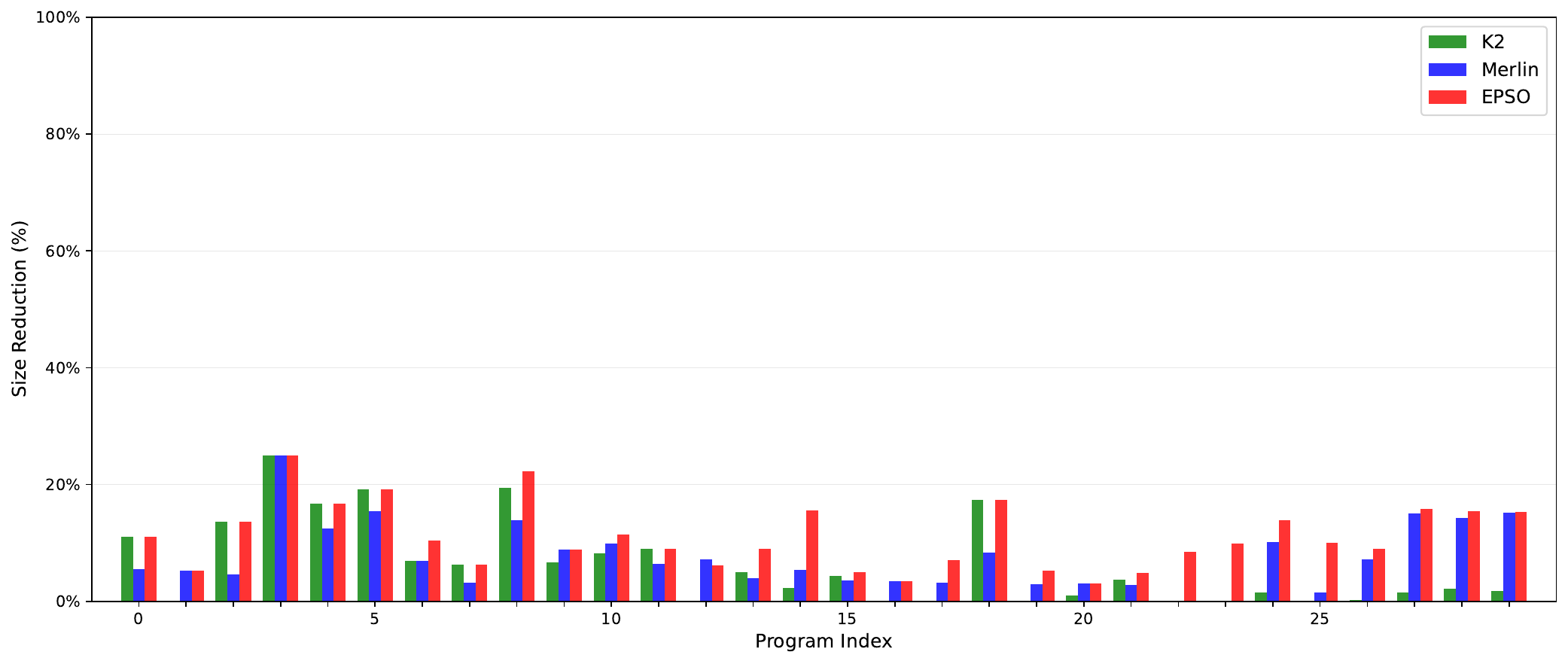}
        \caption{Compactness comparison between EPSO, K2 and Merlin on a diverse benchmark suite.}
        \label{fig:three}
    \end{subfigure}
    
\caption{Comparison of program compactness improvement achieved by EPSO, K2~\cite{xu2021synthesizing} and Merlin\cite{mao2024merlin}across benchmarks from the Linux kernel~\cite{linux_bpf_samples}, Cilium~\cite{cilium}, Katran~\cite{katran}, Sysdig~\cite{sysdig}, Tetragon~\cite{tetragon} and Tracee~\cite{tracee}.}
    \label{fig:size_reduction}
\end{figure*}

\begin{table*}
\center
\caption{Improvements of \textit{estimated} program runtime on benchmarks from Linux kernel and Cilium.}
\scalebox{0.7}{
\begin{tabular}{|l|c|c|c|c|c|c|c|}\hline
\multicolumn{1}{|c|}{\multirow{2}{*}{Benchmark}} & \multicolumn{7}{c|}{Program Runtime(ns)}\\ \cline{2-8}
 & Clang (-Os)  & EPSO  & Compression & K2 & Compression & Merlin & Compression \\ \hline
(1) xdp1\_kern/xdp1 & 25.58 & 23.80 & 6.96\% & 24.55 & 4.03\% & 24.27 & 5.12\% \\ \hline
(2) xdp2\_kern/xdp1 & 30.88 & 29.20 & 5.44\% & 29.76 & 3.63\% & 29.57 & 4.24\% \\ \hline
(3) xdp\_fwd & 51.56 & 41.74 & 19.05\% & 43.73 & 15.19\% & 48.96 & 5.04\% \\ \hline
(4) xdp\_pktcntr & 12.32 & 11.85 & 3.82\% & 11.85 & 3.82\% & 12.21 & 0.89\% \\ \hline
(5) cgroup/recvmsg4 & 65.54 & 65.05 & 0.75\% & 65.46 & 0.11\% & 65.05 & 0.75\% \\ \hline
(6) cgroup/sendmsg4 & 297.37 & 286.75 & 3.57\% & 295.85 & 0.51\% & 297.93 & -0.19\% \\ \hline
Avg. of all benchmarks & & & 6.60\% & & 4.55\% & & 2.64\% \\ \hline
\end{tabular}
}
\label{tab:latency}
\end{table*}

\subsection{RQ1: Program Compactness} 
\label{sec:exp1}

Fig.~\ref{fig:size_reduction} presents a comparison of program compactness improvements achieved by EPSO, K2, and Merlin. In the EPSO vs. K2 comparison (Fig.~\ref{fig:k2}), 
EPSO improves compactness by 6.25\%–25.00\% (avg. 14.40\%) and matches or outperforms K2 on all 19 benchmarks, highlighting the limitations of K2’s stochastic search, which can miss potential optimizations.


\editgroup{
In the EPSO vs. Merlin comparison (Fig.~\ref{fig:sysdig}–Fig.~\ref{fig:tetragon}), EPSO achieves optimization rates ranging from 4.59\% to 68.87\% (avg. 51.30\%) on Sysdig, 0\%–37.79\% (avg. 10.12\%) on Tracee, and 0\%–52.24\% (avg. 13.67\%) on Tetragon. Overall, EPSO outperforms Merlin on 97.28\% of benchmarks on Tracee, 94.82\% on Tetragon, and 86.59\% on Sysdig.} Our analysis shows that 
EPSO performs worse on some samples mainly due to large optimization units that exceed the 10-minute synthesis timeout. To address this, we plan an adaptive timeout strategy that adjusts limits based on program traits, enabling discovery of more complex rewrite rules.


In the comprehensive comparison (Fig.~\ref{fig:three}), EPSO outperforms K2 on all benchmarks and Merlin on 29 of 30 benchmarks. EPSO achieves program compactness improvements from 3.05\% to 25.00\% (avg. 11.12\%), outperforming K2 and Merlin by 82.30\% and 48.46\%, respectively. The only benchmark where EPSO lags behind Merlin involves inter-block optimization, which EPSO does not yet support.

Regarding the comparison between K2 and Merlin, each excels in different scenarios. 
K2 often outperforms Merlin on smaller programs through cost-driven search, but its stochastic nature can miss optimization opportunities. Its performance is also sensitive to parameters. In our experiments, we evaluated 16 different parameter settings and reported the best results.


\begin{tcolorbox}
\textbf{RQ1:} \edit{EPSO achieves up to 68.87\% (avg. 24.37\%) program compactness improvement, outperforming K2 on all benchmarks and Merlin on 92.68\% of them.} It may underperform Merlin when inter-block optimizations are needed or optimization units exceed synthesis timeouts. 

\end{tcolorbox}

\subsection{RQ2: Program Efficiency}
\label{sec:exp2}


\edit{
In efficiency-oriented optimizations, direct execution of candidate programs to measure performance is infeasible, as most are rejected by the verifier. Following K2’s methodology, we estimate program runtime as the sum of instruction latencies, with each instruction profiled by executing its opcode millions of times and recording the average runtime. In real-world networks, efficiency gains boost throughput and reduce packet latency. Our tool, deployed industrially, shows 5.5\% higher QPS under high load, improving service response speed.
}Table~\ref{tab:latency} shows EPSO’s efficiency improvements on six high-performance benchmarks from the Linux kernel~\cite{linux_bpf_samples} (1–4) and Cilium~\cite{cilium} (5–6), compared with K2 and Merlin.


Experiments show that EPSO achieved optimization rates of 0.75\%–19.05\% (avg. 6.60\%), outperforming K2 and Merlin by 45.05\% and 149.77\%, respectively. While K2 and Merlin match EPSO on some benchmarks, K2’s stochastic search and Merlin’s limited rule coverage lead to gaps on others. Notably, Merlin produced a negative rate (-0.19\%) on \textit{cgroup/sendmsg4}, indicating occasional regressions.


\begin{tcolorbox}
\textbf{RQ2:} EPSO improves program efficiency by 0.75\% to 19.05\% (avg. 6.60\%), outperforming both K2 and Merlin across all six benchmarks. In contrast, K2 and Merlin miss certain optimization opportunities due to reliance on stochastic search methods or limited rule coverage.
\end{tcolorbox}

\begin{figure*}[t]
    \small
    \centering
    \begin{subfigure}[t]{0.32\textwidth}
        \includegraphics[width=\linewidth]{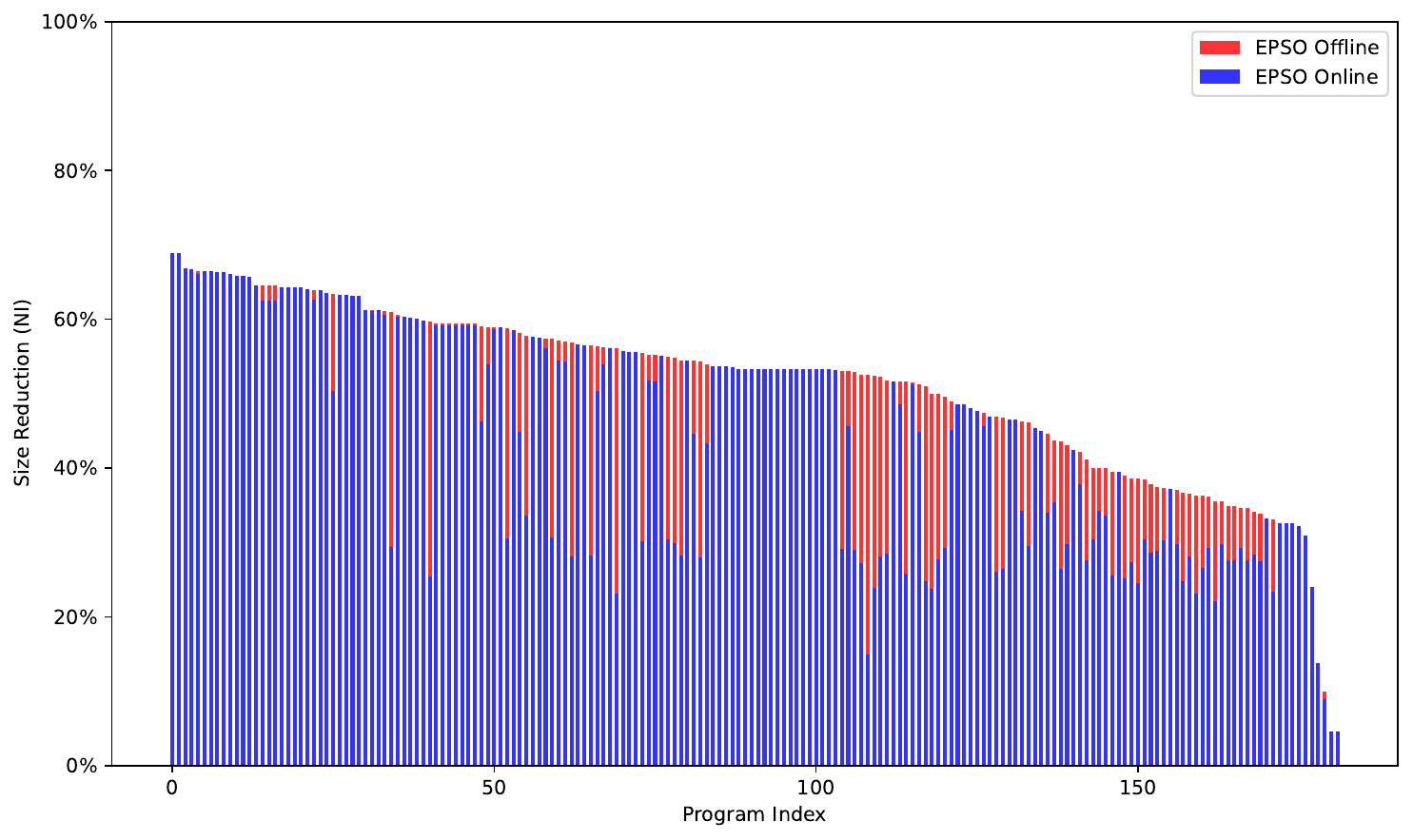}
        \caption{\edit{EPSO compactness on Sysdig~\cite{sysdig}:\\ offline vs. online modes.}}
        \label{fig:sysdig-online}
    \end{subfigure}
    \hfill
    \begin{subfigure}[t]{0.32\textwidth}
        \includegraphics[width=\linewidth]{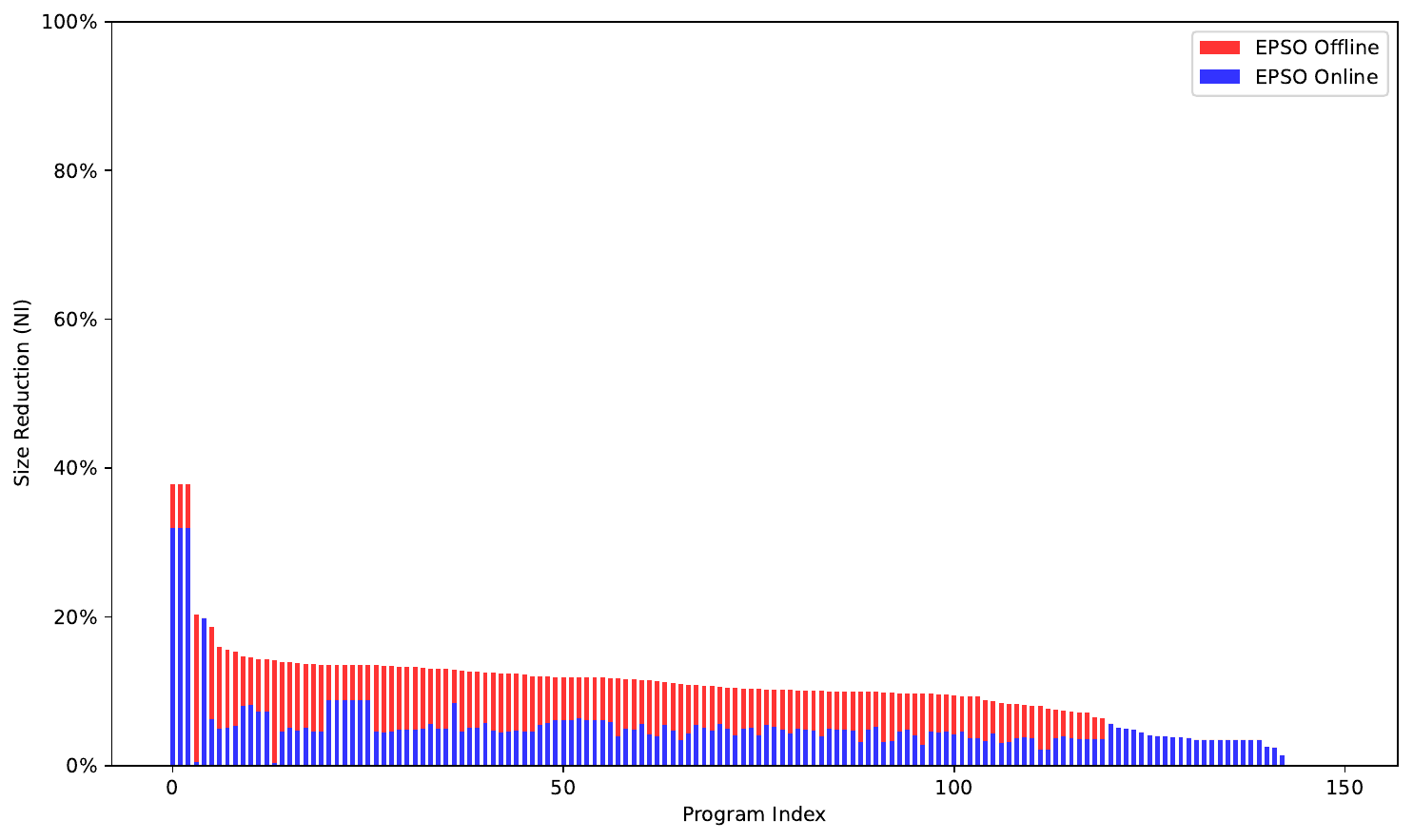}
        \caption{\edit{EPSO compactness on Tracee~\cite{tracee}:\\ offline vs. online modes.}}
        \label{fig:tracee-online}
    \end{subfigure}
    \hfill
    \begin{subfigure}[t]{0.32\textwidth}
        \includegraphics[width=\linewidth]{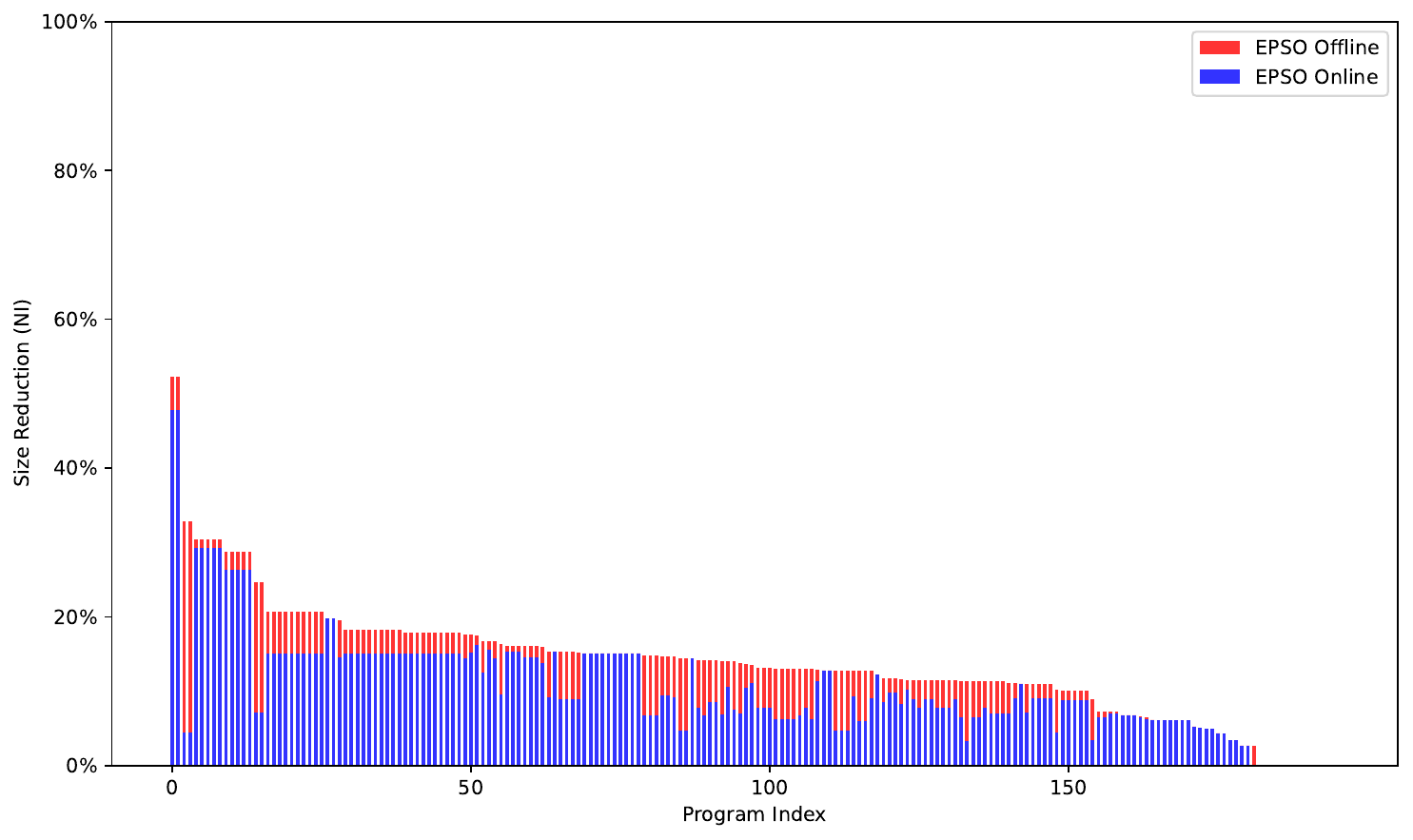}
        \caption{\edit{EPSO compactness on Tetragon~\cite{tetragon}:\\ offline vs. online modes.}}
        \label{fig:tetragon-online}
    \end{subfigure}
    
\caption{\edit{Evaluation of learned rewrite rule effectiveness: offline vs. online.}}
\label{fig:offlin-online-cmp}
\end{figure*}

\subsection{RQ3: Optimization Overhead}
\label{sec:exp3}

We evaluate EPSO's optimization overhead under three configurations: (1) synthesis-only (using only superoptimization); (2) rule-matching-only (using only pre-collected rewrite rules); and (3) hybrid (rule matching first, then synthesis if no rules apply).
For comparison, we also report the optimization overhead of K2, which uses synthesis-based techniques. Merlin, as a rule-based optimizer with negligible runtime overhead, is excluded from this analysis.

\begin{figure}[!t]
\small
    \centering
    \begin{subfigure}{0.23\textwidth}
        \includegraphics[width=\linewidth]{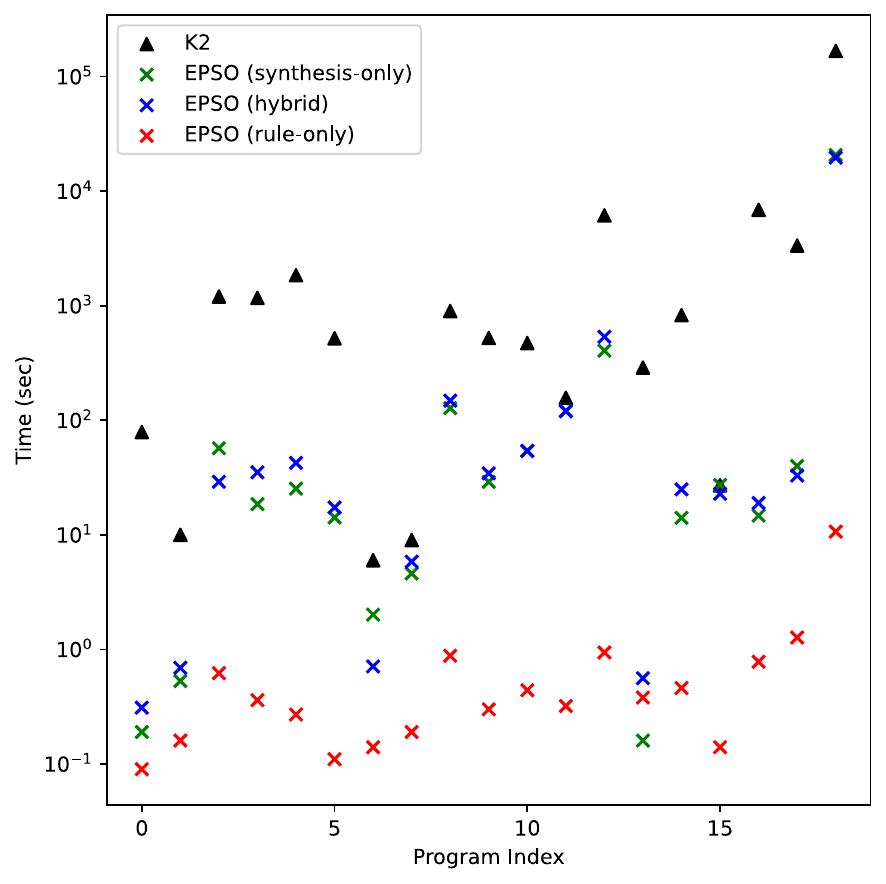}
        \caption{Comparison of optimization overhead between K2 and EPSO (under 3 configurations.)}
        \label{fig:time_cmp}
    \end{subfigure}
    \hfill
    \begin{subfigure}{0.225\textwidth}
        \includegraphics[width=\linewidth]{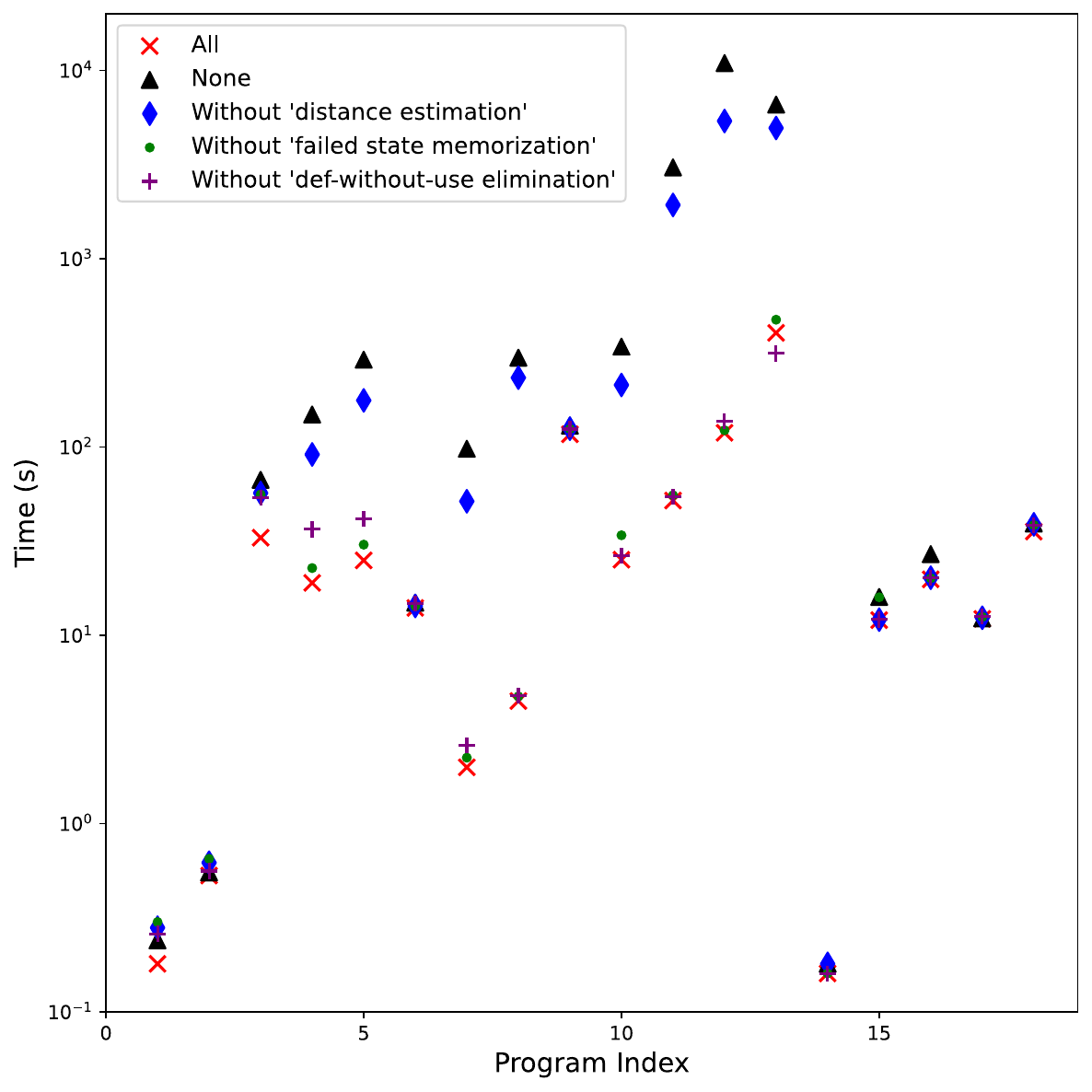}
        \caption{Optimization overhead of EPSO with different pruning strategies applied.}
        \label{fig:time_pruning}
    \end{subfigure}
    \caption{Optimization overhead of K2 and EPSO under different configurations.}
    \label{fig:runtime}
\end{figure}

Fig.~\ref{fig:time_cmp} shows the optimization overhead of K2 and EPSO under three configurations across 19 benchmarks. In synthesis-only mode, EPSO reduces overhead by 88.71\% on average versus K2. In rule-matching-only mode, where rewrite rules are pre-synthesized and reused, EPSO incurs negligible overhead, making it highly effective for large-scale optimizations. The hybrid mode, which prioritizes rule matching and falls back to synthesis, outperforms synthesis-only on some benchmarks but has slightly higher overhead on others due to low rule-match rates.
\edit{Finally, since the rule-extraction and optimization sets used for online optimization are identical, EPSO achieves consistent optimization effects across all three configurations.}


To evaluate how the three pruning strategies reduce optimization overhead, we conducted ablation experiments by disabling each one separately and measuring EPSO’s runtime. 
We also measured overhead when no pruning strategies were used. The results are presented in Fig.~\ref{fig:time_pruning}.

Experiments show that all three pruning strategies significantly reduce optimization overhead, with distance estimation being the most effective. 
Without pruning, synthesis time can increase up to 91×, highlighting pruning’s critical role in improving efficiency.

\begin{tcolorbox}
\textbf{RQ3:} 
EPSO reduces optimization overhead by 88.71\% on average versus K2 in synthesis-only mode, with negligible overhead via rule matching. Pruning, especially distance estimation, cuts synthesis time significantly. Without pruning, synthesis time can increase up to 91×.
\end{tcolorbox}

\subsection{RQ4: Rewrite Rule Effectiveness}
\label{sec:exp4}



\edit{
To evaluate collected rewrite rules, we run superoptimization on 20\% of the Merlin benchmark suite and apply learned rules to the remaining 80\% to test generality. Fig.~\ref{fig:offlin-online-cmp} compares online optimization and direct superoptimization across three projects. Results show rule matching alone reproduces 28.37\%–100\% (avg. 86.69\%), 2.21\%–100\% (avg. 55.63\%), and 13.56\%–100\% (avg. 79.03\%) of superoptimization’s effects on Sysdig, Tracee, and Tetragon, respectively, proving the rules’ strong generality and effectiveness.
}

\editgroup{
During the experiments, we observed that the discovered rewrite rules vary in effectiveness, frequency of application, and ease of discovery. We highlight two representative examples. Example 1 yields significant reductions in instruction count and runtime and is frequently applied, thereby contributing substantially to overall optimization. Example 2 improves register allocation, which is difficult to identify manually but can be automatically uncovered through superoptimization.



\scriptsize
\begin{flushleft} 
\begin{tabular}{ll} 
\textbf{Example1:} & \textbf{Example2:} \\
\texttt{1. r1 = *(u32 *)(r0 + 8)} & \texttt{1. r3 = r7} \\
\texttt{2. r2 = *(u32 *)(r0 + 12)} & \texttt{2. r3 += r1} \\
\texttt{3. r2 <<= 32} & \texttt{3. r3 = *(u16 *)(r3 + 2)} \\
\texttt{4. r2 |= r1} & \texttt{4. r1 = r4} \\
\texttt{=>} & \texttt{=>} \\
\texttt{1. r2 = *(u64 *)(r0 + 8)} & \texttt{1. r1 += r7} \\
\texttt{(r1 is not used afterwards)} & \texttt{2. r3 = *(u16 *)(r1 + 2)} \\
{} & \texttt{3. r1 = r4} \\
\end{tabular}
\end{flushleft}

}

\begin{figure}[t] 
    \centering
    \includegraphics[width=0.6\columnwidth]{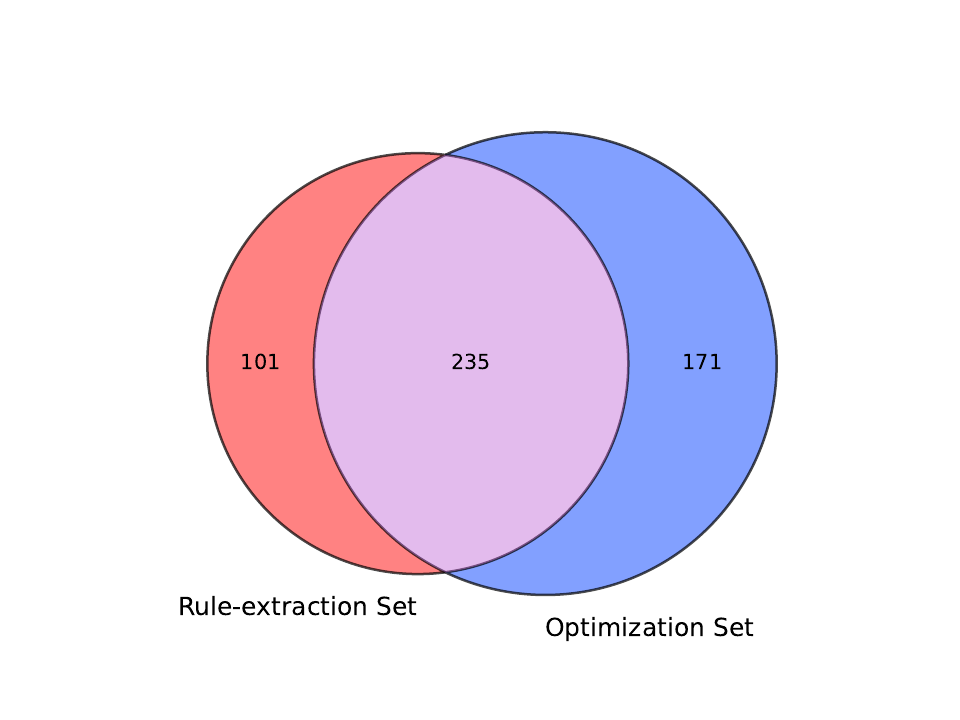} 
    \caption{\edit{Overlap of discovered rewrite rules between rule-extraction and optimization sets.}}
    \label{fig:venn}
\end{figure}

\editgroup{
Fig.~\ref{fig:venn} presents a Venn diagram of rewrite rules from the rule-extraction and optimization sets, where the intersection indicates rules matched during online optimization. 
Despite only 57.88\% of validation-set rewrite rules being discovered during rule extraction and applied in online optimization, frequent use of high-impact rules enables online optimization to achieve, on average, 75.04\% of the effects of direct superoptimization, illustrating variability in rule effectiveness and usage.
}

\begin{tcolorbox}
\textbf{RQ4:} \edit{Evaluating 336 rewrite rules on an unseen benchmark suite, rule matching reproduces 2.21\%–100\% (avg. 75.04\%) of the optimizations achieved by superoptimization, demonstrating the strong generality and effectiveness of the collected rules on unseen programs.}
\end{tcolorbox}

\section{Discussion}

\editgroup{
Prior works K2~\cite{xu2021synthesizing} and Merlin~\cite{mao2024merlin} optimize eBPF programs using synthesis-based and rule-based approaches, respectively, but both exhibit limitations: K2 incurs high optimization overhead, whereas Merlin achieves limited optimization quality. To achieve high-quality optimizations efficiently, our approach adopts a synthesis-based superoptimization strategy like K2, but introduces several key enhancements.


First, we turn superoptimization offline for rewrite rule collection and employ efficient rule matching when optimizing real-world programs, reducing optimization overhead to negligible levels. This is enabled by novel techniques that enhance rule generality and increase successful matches. In contrast, although K2 collects optimization cases during superoptimization, it lacks mechanisms to improve the generality of the collected rules, which is essential for effective reuse. 
Consequently, the narrowly scoped rules rarely match new code, resulting in repeated and costly synthesis and, therefore, high optimization overhead. 
Second, unlike K2’s stochastic search, we employ enumerative search to uncover more optimization opportunities. Coupled with our pruning strategies, synthesis is further accelerated, making EPSO more efficient than K2 even as a pure superoptimizer.
}

While our approach demonstrates efficient and effective eBPF optimization, several extensions merit further exploration. First, supporting inter-block optimization could enable deeper transformations by allowing modifications across basic block boundaries, rather than restricting them to intra-block scopes. 
Second, adaptive adjustment of window size and timeout based on program structure could further reduce overhead and enhance optimization quality. Currently, discovering diverse rewrites requires multiple superoptimization rounds with varying window sizes. \edit{In our experiments, 795 well-generalized rewrite rules were identified with window sizes ranging from 1 to 62.} As synthesis time depends more on rewrite size than window size, EPSO often completes within the timeout even for large programs, provided rewrites are small. However, a fixed timeout may impede the discovery of more complex rewrites that require longer synthesis. Developing adaptive strategies for window sizing and timeout tuning thus represents a promising direction to further improve both efficiency and optimization quality.

\section{Related Work}

Program synthesis~\cite{gulwani2017program} aims to generate programs that satisfy given specifications, such as input-output examples, demonstrations, or natural language descriptions. In the context of code optimization, the specification is the original program, and the goal is to produce functionally equivalent code with improved performance.



Superoptimization~\cite{massalin1987superoptimizer} leverages program synthesis to find more efficient equivalents by exhaustively exploring the program space. Common strategies include enumerative search~\cite{massalin1987superoptimizer,granlund1992eliminating,bansal2006automatic,phothilimthana2016scaling,alur2017scaling,huang2020reconciling,alur2015synthesis,srinivasan2015synthesis}, constraint solving~\cite{joshi2002denali,jangda2017unbounded,jeon2015adaptive,gulwani2011synthesis,reynolds2015counterexample,jha2010oracle}, and stochastic search~\cite{schkufza2013stochastic,xu2021synthesizing}. Although superoptimization produces high-quality results, its practicality is often limited by high computational cost, especially for large-scale programs. Here, we adopt enumerative search with pruning for efficient optimizations. In addition to established pruning techniques~\cite{massalin1987superoptimizer,granlund1992eliminating,phothilimthana2016scaling}, we also introduce a novel distance-estimation-based pruning strategy tailored for eBPF’s register-centric semantics, which is highly effective in the BPF context.

Despite various pruning strategies, program synthesis remains time-consuming due to its exhaustive search nature and scales poorly to large programs. Inspired by prior work~\cite{bansal2006automatic,liu2024minotaur}, we offload superoptimization to an offline phase to extract high-quality rewrite rules, which are then efficiently applied to real-world BPF programs via rule matching. To enhance rule generality and reusability, we design a specialized slicing strategy and instruction abstraction mechanism tailored for BPF instructions, both proven effective.


\section{Conclusion}
We propose an optimization approach for eBPF programs that shifts the costly superoptimization process to an offline phase for discovering rewrite rules, while employing lightweight rule matching to optimize real-world BPF programs. This approach effectively combines the strengths of synthesis-based and rule-based techniques, achieving high-quality optimizations with negligible runtime overhead. \edit{Across diverse BPF programs of varying sources and scales, EPSO discovers a total of 795 rewrite rules. It achieves up to 68.87\% (avg. 24.37\%) improvement in program compactness compared to Clang’s output, outperforming the state-of-the-art BPF optimizer K2 on all benchmarks and Merlin on 92.68\% of them.} Furthermore, EPSO reduces program runtime by an average of 6.60\%, thereby enhancing throughput and lowering latency in network applications. Additionally, it reduces optimization overhead by 88.71\% compared to K2 in synthesis-only mode, while incurring negligible overhead when employing rule matching.

\section*{Acknowledgments}

We are grateful for the constructive feedback of all anonymous reviewers to improve this manuscript. The authors are supported in part by the National Key Research and Development Program of China (2024YFB2505604) and the National Natural Science Foundation of China (No.62232008, 62172200).

\clearpage
\balance
\bibliographystyle{IEEEtran}
\bibliography{main}

@inproceedings{alur2015synthesis,
  title={Synthesis through unification},
  author={Alur, Rajeev and {\v{C}}ern{\`y}, Pavol and Radhakrishna, Arjun},
  booktitle={International Conference on Computer Aided Verification},
  pages={163--179},
  year={2015},
  organization={Springer}
}

@inproceedings{srinivasan2015synthesis,
  title={Synthesis of machine code from semantics},
  author={Srinivasan, Venkatesh and Reps, Thomas},
  booktitle={Proceedings of the 36th ACM SIGPLAN Conference on Programming Language Design and Implementation},
  pages={596--607},
  year={2015}
}

@inproceedings{reynolds2015counterexample,
  title={Counterexample-guided quantifier instantiation for synthesis in SMT},
  author={Reynolds, Andrew and Deters, Morgan and Kuncak, Viktor and Tinelli, Cesare and Barrett, Clark},
  booktitle={Computer Aided Verification: 27th International Conference, CAV 2015, San Francisco, CA, USA, July 18-24, 2015, Proceedings, Part II 27},
  pages={198--216},
  year={2015},
  organization={Springer}
}

@article{gulwani2011synthesis,
  title={Synthesis of loop-free programs},
  author={Gulwani, Sumit and Jha, Susmit and Tiwari, Ashish and Venkatesan, Ramarathnam},
  journal={ACM SIGPLAN Notices},
  volume={46},
  number={6},
  pages={62--73},
  year={2011},
  publisher={ACM New York, NY, USA}
}

@inproceedings{jeon2015adaptive,
  title={Adaptive concretization for parallel program synthesis},
  author={Jeon, Jinseong and Qiu, Xiaokang and Solar-Lezama, Armando and Foster, Jeffrey S},
  booktitle={International Conference on Computer Aided Verification},
  pages={377--394},
  year={2015},
  organization={Springer}
}

@inproceedings{huang2020reconciling,
  title={Reconciling enumerative and deductive program synthesis},
  author={Huang, Kangjing and Qiu, Xiaokang and Shen, Peiyuan and Wang, Yanjun},
  booktitle={Proceedings of the 41st ACM SIGPLAN Conference on Programming Language Design and Implementation},
  pages={1159--1174},
  year={2020}
}

@article{liu2024minotaur,
  title={Minotaur: A SIMD-oriented synthesizing superoptimizer},
  author={Liu, Zhengyang and Mada, Stefan and Regehr, John},
  journal={Proceedings of the ACM on Programming Languages},
  volume={8},
  number={OOPSLA2},
  pages={1561--1585},
  year={2024},
  publisher={ACM New York, NY, USA}
}

@misc{cilium,
  year = {2025},
  url = {https://github.com/cilium/cilium},
  title = {Cilium},
  note = {Accessed: 2025-03-19},
}

@misc{katran,
  year = {2025},
  url = {https://github.com/facebookincubator/katran},
  title = {Katran},
  note = {Accessed: 2025-03-19},
}

@misc{tetragon,
  year = {2025},
  url = {https://github.com/cilium/tetragon},
  title = {Tetragon},
  note = {Accessed: 2025-03-19},
}

@misc{bpftrace,
  year = {2025},
  url = {https://github.com/bpftrace/bpftrace},
  title = {bpftrace},
  note = {Accessed: 2025-04-27},
}

@misc{linux_bpf_samples,
  year = {2025},
  url = {https://github.com/torvalds/linux/tree/master/samples/bpf},
  title = {BPF samples from Linux},
  note = {Accessed: 2025-03-19},
}

@article{korf1985depth,
  title={Depth-first iterative-deepening: An optimal admissible tree search},
  author={Korf, Richard E},
  journal={Artificial intelligence},
  volume={27},
  number={1},
  pages={97--109},
  year={1985},
  publisher={Elsevier}
}

@misc{sysdig,
  year = {2025},
  url = {https://github.com/draios/sysdig},
  title = {Sysdig},
  note = {Accessed: 2025-03-19},
}

@misc{tracee,
  year = {2025},
  url = {https://github.com/aquasecurity/tracee},
  title = {Tracee},
  note = {Accessed: 2025-03-19},
}

@article{brunella2022hxdp,
  title={hXDP: Efficient software packet processing on FPGA NICs},
  author={Brunella, Marco Spaziani and Belocchi, Giacomo and Bonola, Marco and Pontarelli, Salvatore and Siracusano, Giuseppe and Bianchi, Giuseppe and Cammarano, Aniello and Palumbo, Alessandro and Petrucci, Luca and Bifulco, Roberto},
  journal={Communications of the ACM},
  volume={65},
  number={8},
  pages={92--100},
  year={2022},
  publisher={ACM New York, NY, USA}
}

@book{solar2008program,
  title={Program synthesis by sketching},
  author={Solar-Lezama, Armando},
  year={2008},
  publisher={University of California, Berkeley}
}

@inproceedings{bonola2022faster,
  title={Faster Software Packet Processing on $\{$FPGA$\}$$\{$NICs$\}$ with $\{$eBPF$\}$ Program Warping},
  author={Bonola, Marco and Belocchi, Giacomo and Tulumello, Angelo and Brunella, Marco Spaziani and Siracusano, Giuseppe and Bianchi, Giuseppe and Bifulco, Roberto},
  booktitle={2022 USENIX Annual Technical Conference (USENIX ATC 22)},
  pages={987--1004},
  year={2022}
}

@inproceedings{kuo2022verified,
  title={Verified programs can party: optimizing kernel extensions via post-verification merging},
  author={Kuo, Hsuan-Chi and Chen, Kai-Hsun and Lu, Yicheng and Williams, Dan and Mohan, Sibin and Xu, Tianyin},
  booktitle={Proceedings of the Seventeenth European Conference on Computer Systems},
  pages={283--299},
  year={2022}
}

@misc{eBPFDocumentary,
  year = {2024},
  url = {https://ebpf.io/what-is-ebpf/},
  title = {eBPF Documentary},
  note = {Accessed: 2024-06-01},
}

@inproceedings{mccanne1993bsd,
  title={The BSD Packet Filter: A New Architecture for User-level Packet Capture.},
  author={McCanne, Steven and Jacobson, Van},
  booktitle={USENIX winter},
  volume={46},
  year={1993}
}

@inproceedings{hoiland2018express,
  title={The express data path: Fast programmable packet processing in the operating system kernel},
  author={H{\o}iland-J{\o}rgensen, Toke and Brouer, Jesper Dangaard and Borkmann, Daniel and Fastabend, John and Herbert, Tom and Ahern, David and Miller, David},
  booktitle={Proceedings of the 14th international conference on emerging networking experiments and technologies},
  pages={54--66},
  year={2018}
}

@article{vieira2020fast,
  title={Fast packet processing with ebpf and xdp: Concepts, code, challenges, and applications},
  author={Vieira, Marcos AM and Castanho, Matheus S and Pac{\'\i}fico, Racyus DG and Santos, Elerson RS and J{\'u}nior, Eduardo PM C{\^a}mara and Vieira, Luiz FM},
  journal={ACM Computing Surveys (CSUR)},
  volume={53},
  number={1},
  pages={1--36},
  year={2020},
  publisher={ACM New York, NY, USA}
}

@article{miano2023fast,
  title={Fast in-kernel traffic sketching in EBPF},
  author={Miano, Sebastiano and Chen, Xiaoqi and Basat, Ran Ben and Antichi, Gianni},
  journal={ACM SIGCOMM Computer Communication Review},
  volume={53},
  number={1},
  pages={3--13},
  year={2023},
  publisher={ACM New York, NY, USA}
}

@inproceedings{scholz2018performance,
  title={Performance implications of packet filtering with linux ebpf},
  author={Scholz, Dominik and Raumer, Daniel and Emmerich, Paul and Kurtz, Alexander and Lesiak, Krzysztof and Carle, Georg},
  booktitle={2018 30th International Teletraffic Congress (ITC 30)},
  volume={1},
  pages={209--217},
  year={2018},
  organization={IEEE}
}

@article{caviglione2021kernel,
  title={Kernel-level tracing for detecting stegomalware and covert channels in Linux environments},
  author={Caviglione, Luca and Mazurczyk, Wojciech and Repetto, Matteo and Schaffhauser, Andreas and Zuppelli, Marco},
  journal={Computer Networks},
  volume={191},
  pages={108010},
  year={2021},
  publisher={Elsevier}
}

@book{calavera2019linux,
  title={Linux Observability with BPF: Advanced Programming for Performance Analysis and Networking},
  author={Calavera, David and Fontana, Lorenzo},
  year={2019},
  publisher={O'Reilly Media}
}

@book{gregg2019bpf,
  title={BPF performance tools},
  author={Gregg, Brendan},
  year={2019},
  publisher={Addison-Wesley Professional}
}

@article{soldani2023ebpf,
  title={ebpf: A new approach to cloud-native observability, networking and security for current (5g) and future mobile networks (6g and beyond)},
  author={Soldani, David and Nahi, Petrit and Bour, Hami and Jafarizadeh, Saber and Soliman, Mohammed F and Di Giovanna, Leonardo and Monaco, Francesco and Ognibene, Giuseppe and Risso, Fulvio},
  journal={IEEE Access},
  volume={11},
  pages={57174--57202},
  year={2023},
  publisher={IEEE}
}

@misc{security2019blog,
  author = {Sean Kerner},
  year = {2019},
  url = {https://lwn.net/Articles/790684/},
  title = {BPF for security—and chaos—in Kubernetes},
  note = {Accessed:2024-06-01}
}

@article{wang2022design,
  title={Design and implementation of an intrusion detection system by using Extended BPF in the Linux kernel},
  author={Wang, Shie-Yuan and Chang, Jen-Chieh},
  journal={Journal of Network and Computer Applications},
  volume={198},
  pages={103283},
  year={2022},
  publisher={Elsevier}
}

@article{nam2022secure,
  title={Secure inter-container communications using XDP/eBPF},
  author={Nam, Jaehyun and Lee, Seungsoo and Porras, Phillip and Yegneswaran, Vinod and Shin, Seungwon},
  journal={IEEE/ACM Transactions on Networking},
  volume={31},
  number={2},
  pages={934--947},
  year={2022},
  publisher={IEEE}
}

@inproceedings{xu2021synthesizing,
  title={Synthesizing safe and efficient kernel extensions for packet processing},
  author={Xu, Qiongwen and Wong, Michael D and Wagle, Tanvi and Narayana, Srinivas and Sivaraman, Anirudh},
  booktitle={Proceedings of the 2021 ACM SIGCOMM 2021 Conference},
  pages={50--64},
  year={2021}
}

@inproceedings{mao2024merlin,
  title={Merlin: Multi-tier Optimization of eBPF Code for Performance and Compactness},
  author={Mao, Jinsong and Ding, Hailun and Zhai, Juan and Ma, Shiqing},
  booktitle={Proceedings of the 29th ACM International Conference on Architectural Support for Programming Languages and Operating Systems, Volume 3},
  pages={639--653},
  year={2024}
}

@article{gulwani2017program,
  title={Program synthesis},
  author={Gulwani, Sumit and Polozov, Oleksandr and Singh, Rishabh and others},
  journal={Foundations and Trends{\textregistered} in Programming Languages},
  volume={4},
  number={1-2},
  pages={1--119},
  year={2017},
  publisher={Now Publishers, Inc.}
}

@article{massalin1987superoptimizer,
  title={Superoptimizer: a look at the smallest program},
  author={Massalin, Henry},
  journal={ACM SIGARCH Computer Architecture News},
  volume={15},
  number={5},
  pages={122--126},
  year={1987},
  publisher={ACM New York, NY, USA}
}

@inproceedings{granlund1992eliminating,
  title={Eliminating branches using a superoptimizer and the GNU C compiler},
  author={Granlund, Torbj{\"o}rn and Kenner, Richard},
  booktitle={Proceedings of the ACM SIGPLAN 1992 conference on Programming language design and implementation},
  pages={341--352},
  year={1992}
}

@article{bansal2006automatic,
  title={Automatic generation of peephole superoptimizers},
  author={Bansal, Sorav and Aiken, Alex},
  journal={ACM SIGARCH Computer Architecture News},
  volume={34},
  number={5},
  pages={394--403},
  year={2006},
  publisher={ACM New York, NY, USA}
}

@inproceedings{alur2017scaling,
  title={Scaling enumerative program synthesis via divide and conquer},
  author={Alur, Rajeev and Radhakrishna, Arjun and Udupa, Abhishek},
  booktitle={International conference on tools and algorithms for the construction and analysis of systems},
  pages={319--336},
  year={2017},
  organization={Springer}
}

@inproceedings{phothilimthana2016scaling,
  title={Scaling up superoptimization},
  author={Phothilimthana, Phitchaya Mangpo and Thakur, Aditya and Bodik, Rastislav and Dhurjati, Dinakar},
  booktitle={Proceedings of the Twenty-First International Conference on Architectural Support for Programming Languages and Operating Systems},
  pages={297--310},
  year={2016}
}

@article{joshi2002denali,
  title={Denali: A goal-directed superoptimizer},
  author={Joshi, Rajeev and Nelson, Greg and Randall, Keith},
  journal={ACM SIGPLAN Notices},
  volume={37},
  number={5},
  pages={304--314},
  year={2002},
  publisher={ACM New York, NY, USA}
}

@inproceedings{jha2010oracle,
  title={Oracle-guided component-based program synthesis},
  author={Jha, Susmit and Gulwani, Sumit and Seshia, Sanjit A and Tiwari, Ashish},
  booktitle={Proceedings of the 32nd ACM/IEEE International Conference on Software Engineering-Volume 1},
  pages={215--224},
  year={2010}
}

@inproceedings{jangda2017unbounded,
  title={Unbounded superoptimization},
  author={Jangda, Abhinav and Yorsh, Greta},
  booktitle={Proceedings of the 2017 ACM SIGPLAN International Symposium on New Ideas, New Paradigms, and Reflections on Programming and Software},
  pages={78--88},
  year={2017}
}

@article{schkufza2013stochastic,
  title={Stochastic superoptimization},
  author={Schkufza, Eric and Sharma, Rahul and Aiken, Alex},
  journal={ACM SIGARCH Computer Architecture News},
  volume={41},
  number={1},
  pages={305--316},
  year={2013},
  publisher={ACM New York, NY, USA}
}

\end{document}